\input ./style/arxiv-general.cfg
\documentclass[aoas,MSNbibl,nameyear,seceqn,dvips]{arximspdf}
\makeatletter
   \@ifpackageloaded{graphicx}{}{\usepackage{graphicx}}
\makeatother
\usepackage{algorithm,algorithmic}
\usepackage{url,breakurl}

\doi{10.1214/15-AOAS842}
\volume{9}
\issue{3}
\pubyear{2015}
\firstpage{1103}
\lastpage{1140}
\docsubty{FLA}

\makeatletter
\newcommand{\rrVert}{\Vert}
\newcommand{\llVert}{\Vert}
\newtheorem{theorem}{Theorem}[section]
\newtheorem{lemma}[theorem]{Lemma}
\newtheorem{proposition}[theorem]{Proposition}
\newproclaim{assumption}[theorem]{Assumption}
\newcommand{\eqref}[1]{(\ref{#1})}
\newcommand{\R}{\mathbb{R}}
\newcommand{\LBH}{\lambda_{\mathrm{BH}}}
\newcommand{\LG}{\lambda_{\mathrm{G}}}
\newcommand{\LCV}{\lambda_{\mathrm{CV}}}
\newcommand{\LMC}{\lambda_{\mathrm{MC}}}
\newcommand{\eqdef}{\stackrel{\mathrm{def}}{=}}
\newcommand{\iBH}{i_{\mathrm{BH}}}
\newcommand{\proxh}{\operatorname{prox}}
\newcommand{\LBONF}{\lambda_{\mathrm{Bonf}}}
\newcommand{\betaS}{\beta_{\mathcal{S}}}
\newcommand{\XS}{X_{\mathcal{S}}}
\newcommand{\lambdaS}{\lambda_{\mathcal{S}}}
\makeatother

\begin{document}
\begin{frontmatter}

\title{SLOPE---Adaptive variable selection via convex~optimization}
\runtitle{Sorted L-One Penalized Estimation}

\begin{aug}
\author[A]{\fnms{Ma{\l}gorzata}~\snm{Bogdan}\thanksref{M1,T1}},
\author[B]{\fnms{Ewout}~\snm{van den Berg}\thanksref{M2,T2}},
\author[C]{\fnms{Chiara}~\snm{Sabatti}\thanksref{M3,T3}},
\author[D]{\fnms{Weijie}~\snm{Su}\thanksref{M3,T4}}
\and
\author[E]{\fnms{Emmanuel J.}~\snm{Cand\`es}\corref{}\ead[label=e1]{candes@stanford.edu}\thanksref{M3,T5}}
\runauthor{M. Bogdan et al.}
\affiliation{Wroc{\l}aw University of Technology\thanksmark{M1},
IBM T.J. Watson Research Center\thanksmark{M2}\\ and
Stanford University\thanksmark{M3}}
\address[A]{M. Bogdan\\
Department of Mathematics\\
Wroc{\l}aw University of Technology\\
50-370 Wroc{\l}aw\\
Poland}
\address[B]{E. van den Berg\\
Human Language Technologies\\
IBM T.J. Watson Research Center\\
Yorktown Heights, New York 10598\\
USA}
\address[C]{C. Sabatti\\
Department of Health Research and Policy\\
Division of Biostatistics\\
Stanford University\\
HRP Redwood Building\\
Stanford, California 94305\\
USA\\
and\\
Department of Statistics\\
Stanford University\\
390 Serra Mall, Sequoia Hall\\
Stanford, California 94305\\
USA}
\address[D]{W. Su\\
Department of Statistics\\
Stanford University\\
90 Serra Mall, Sequoia Hall\\
Stanford, California 94305\hspace*{32.5pt}\\
USA}
\address[E]{E. J. Cand\`es\\
Department of Statistics\\
Stanford University\\
390 Serra Mall, Sequoia Hall\\
Stanford, California 94305\\
USA\\
and\\
Department of Mathematics\\
Stanford University\\
Building 380\\
Stanford, California 94305\\
USA\\
\printead{e1}}
\end{aug}
\thankstext{T1}{Supported in part by a Fulbright Scholarship, NSF Grant DMS-10-43204 and the European Union's 7th Framework Programme for
research, technological development and demonstration under Grant
Agreement no. 602552.}
\thankstext{T2}{Supported in part by NSF Grant
DMS-09-06812 (American Reinvestment and Recovery Act).}
\thankstext{T3}{Supported in part by NIH Grants HG006695 and MH101782.}
\thankstext{T4}{Supported in part by a General Wang Yaowu Stanford
Graduate Fellowship.}
\thankstext{T5}{Supported in part by AFOSR under Grant
FA9550-09-1-0643, by ONR under Grant N00014-09-1-0258 and by a gift
from the Broadcom Foundation.}

%
\received{\smonth{5} \syear{2014}}
%
\revised{\smonth{2} \syear{2015}}

%
\begin{abstract}
We introduce a new estimator for the vector of coefficients $\beta$
in the linear model $y = X \beta+ z$, where $X$ has dimensions
$n\times p$ with $p$ possibly larger than~$n$. SLOPE, short for
Sorted L-One Penalized Estimation, is the solution to
\[
\min_{b \in\mathbb{R}^p} \frac{1}{2} \llVert y - Xb\rrVert
_{\ell_2}^2 + \lambda_1 \vert b\vert
_{(1)} + \lambda_2 \vert b\vert_{(2)} + \cdots+
\lambda_p \vert b\vert_{(p)},
\]
where $\lambda_1 \ge\lambda_2 \ge\cdots\ge\lambda_p \ge0$ and
$\vert b\vert_{(1)} \ge\vert b\vert_{(2)} \ge\cdots\ge\vert
b\vert_{(p)}$ are the decreasing
absolute values of the entries of $b$. This is a convex program and
we demonstrate a solution algorithm whose computational complexity is
roughly comparable to that of classical $\ell_1$ procedures such as
the Lasso.
Here, the regularizer is a sorted $\ell_1$ norm, which
penalizes the regression coefficients according to their rank: the
higher the rank---that is, stronger the signal---the larger the
penalty. This is similar to the
Benjamini and Hochberg [\textit{J.~Roy. Statist. Soc. Ser. B} \textbf{57} (1995) 289--300]
procedure (BH) which
compares more significant $p$-values with more
stringent thresholds. One notable choice of the sequence
$\{\lambda_i\}$ is given by the BH critical values $\lambda_{\mathrm
{BH}} (i) = z(1 - i
\cdot q/2p)$, where $q\in(0,1)$ and $z(\alpha)$ is the quantile of a
standard normal distribution. SLOPE aims to provide finite sample
guarantees on the selected model; of special interest is the false
discovery rate (FDR), defined as the expected proportion of irrelevant
regressors among all selected predictors. Under orthogonal designs,
SLOPE with $\lambda_{\mathrm{BH}}$ provably controls FDR at level
$q$. Moreover, it
also appears to have appreciable inferential properties under more
general designs $X$ while having substantial power, as demonstrated in
a series of experiments running on both simulated and real data.
\end{abstract}

%
\end{frontmatter}

\setcounter{footnote}{5}

\section*{Introduction}
Analyzing and extracting information from data sets where the number of
observations $n$ is smaller than the number of variables $p$ is one of
the challenges of the present ``big-data'' world. In response, the
statistics literature of the past two decades documents the
development of a variety of methodological approaches to address this
challenge. A frequently discussed problem is that of linking, through
a linear model, a response variable $y$ to a set of predictors
$\{X_j\}$ taken from a very large family of possible explanatory
variables. In this context, the Lasso [\citet{Tibs96}] and the Dantzig
selector [\citet{DS}], for example, are computationally attractive
procedures offering some theoretical guarantees, and with consequent
widespread application. In spite of this, there are some scientific
problems where the outcome of these procedures is not entirely
satisfying, as they do not come with a machinery allowing us to make
inferential statements on the validity of selected models in finite
samples. To illustrate this, we resort to an example.
%

Consider a study where a geneticist has collected information about
$n$ individuals by having identified and measured all $p$ possible
genetics variants in a genomic region. The geneticist wishes to
discover which variants cause a certain biological phenomenon, such as an
increase in blood cholesterol level.
Measuring cholesterol levels in a new individual is cheaper and faster
than scoring his or her genetic variants, so that predicting $y$ in
future samples given the value of the relevant covariates is not an
important goal. Instead, correctly identifying functional variants is
relevant. A genetic polymorphism \textit{correctly}
implicated in the determination of cholesterol levels points to a
specific gene and to a biological pathway that might not be previously
known to be related to blood lipid levels and, therefore, promotes an
increase in our understanding of biological mechanisms, as well as
providing targets for drug development. On the other hand, the
\textit{erroneous} discovery of an association between a genetic
variant and
cholesterol levels will translate to a considerable waste of time and
money, which will be spent in trying to verify this association with
direct manipulation experiments.
It is worth emphasizing
that some of the genetic variants in the study have a biological
effect while others do not---there is a ground truth that
statisticians can aim to discover.
To be able to share the
results with the scientific community in a convincing manner, the
researcher needs to be able to attach some finite sample confidence
statements to his/her findings. In a more abstract language, our
geneticist would need a tool that privileges correct model selection
over minimization of prediction error, and would allow for inferential
statements to be made on the validity of his/her selections. This paper
presents a new methodology that attempts to address some of these
needs. 

We imagine that the $n$-dimensional response vector $y$ is truly
generated by a linear model of the form
\[
\label{eq:linear} y = X \beta+ z,
\]
with $X$ an $n \times p$ design matrix, $\beta$ a $p$-dimensional
vector of regression coefficients and $z$ an $n \times1$ vector of
random errors. We assume that all relevant variables (those with
$\beta_i\neq0$) are measured in addition to a large number of
irrelevant ones. As any statistician knows, these assumptions are quite
restrictive, but they are a widely accepted starting
point. To formalize our goal, namely, the selection of important
variables accompanied by a finite sample confidence statement, we seek
a procedure that controls the expected proportion of irrelevant
variables among the selected. In a scientific context where selecting
a variable corresponds to making a discovery, we aim at controlling
the False Discovery Rate (FDR). The FDR is of course a well-recognized
measure of global error in multiple testing and effective procedures
to control it are available: indeed, the \citet{BH95}
procedure (BH) inspired the present proposal.
The connection between multiple testing and model
selection has been made before [see, e.g., \citet
{Bauer,RIC,AbramovichBenjamini95,ABDJ} and
\citet{QREI}] and others in recent literature have tackled the challenges
encountered by our geneticists: we will discuss the differences
between our approach and others in later sections as appropriate. The
procedure we introduce in this paper is, however, entirely
new. Variable selection is achieved by solving a convex problem not
previously considered in the statistical literature, and which marries
the advantages of $\ell_1$ penalization with the adaptivity inherent
in strategies like BH.

%
Section~\ref{sec1} of this paper introduces SLOPE, our novel penalization
strategy, motivates its
construction in the context of orthogonal designs, and places it in
the context of current knowledge of effective model selection
strategies. Section~\ref{sec2} describes the algorithm we developed and
implemented to find SLOPE estimates. Section~\ref{sec3} showcases the
application of our novel procedure in a variety of settings: we
illustrate how it effectively solves a multiple testing problem with
positively correlated test statistics; we discuss how regularizing
parameters should be chosen in nonorthogonal designs; we investigate
the robustness of SLOPE to some
violations of model assumptions and we apply it
to a genetic data set, not unlike our idealized example. Section~\ref{sec4}
concludes the paper with a discussion comparing our methodology to
other recently introduced proposals as well as outlining open
problems.

\section{Sorted L-One Penalized Estimation (SLOPE)}\label{sec1}
\label{sec:slope}


\subsection{Adaptive penalization and multiple testing in orthogonal
designs}\label{sec1.1}

To build intuition behind SLOPE, which encompasses our proposal for
model selection in situations where $p>n$, we begin by considering the
case of orthogonal designs and i.i.d.~Gaussian errors with known
standard deviation, as this makes the connection between model
selection and multiple testing natural. Since the design is
orthogonal, $X'X=I_p$, and the regression $y =X\beta+z$ with $z \sim
\mathcal{N}(0, \sigma^2 I_n)$ can be recast as
%
%
\begin{equation}
\label{multtest} \tilde y = X' y=X'X\beta+X'
z=\beta+ X' z \sim\mathcal{N} \bigl(\beta, \sigma^2
I_p \bigr).
\end{equation}
In some sense, the problem of selecting the correct model reduces to
the problem of testing the $p$ hypotheses $H_{0,j}: \beta_j = 0$
versus two-sided alternatives $H_{1,j}: \beta_i \neq0$. When $p$ is
large, a multiple comparison correction strategy is called for and we
consider two popular procedures:
\begin{itemize}
\item{\em Bonferroni's method.} To control the familywise error
rate\footnote{Recall that the FWER is the probability of at least
one false rejection.} (FWER) at level $\alpha\in[0,1]$, one can
apply Bonferroni's method, and reject $H_{0,j}$ if $\vert\tilde
y_j\vert/\sigma> \Phi^{-1}(1-\alpha/2p)$, where $\Phi^{-1}(\alpha
)$ is
the $\alpha$th quantile of the standard normal distribution. Hence,
Bonferroni's method defines a comparison threshold that depends only
on the number of covariates, $p$, and the noise level.
\item{\em Benjamini--Hochberg step-up procedure.} To control the
FDR at level $q\in[0,1]$, BH begins by sorting the entries of
$\tilde y$ in decreasing order of magnitude, $\vert\tilde y\vert
_{(1)} \ge
\vert\tilde y\vert_{(2)} \ge\cdots\ge\vert\tilde y\vert
_{(p)}$, which yields
corresponding ordered hypotheses $H_{(1)}, \ldots, H_{(p)}$. [Note
that here, as in the rest of the paper, $(1)$ indicates the largest
element of a set, instead of the smallest. This breaking with common
convention allows us to keep (1) as the index for the most
``interesting'' hypothesis]. Then BH rejects all hypotheses $H_{(i)}$ for
which $i \le i_{\mathrm{BH}}$, where $i_{\mathrm{BH}}$ is defined by
%
%
\begin{equation}
\label{eq:stepup} i_{\mathrm{BH}} = \max\bigl\{i: \vert\tilde y\vert
_{(i)}/\sigma\geq\Phi^{-1}(1- q_i) \bigr\}, \qquad
q_i = i \cdot q/2p
\end{equation}
(with the convention that $i_{\mathrm{BH}} = 0$ if the set above is
empty).
Letting $V$
(resp.,~$R$) be the total number of false rejections (resp., total
number of rejections), \citet{BH95} showed that for BH
%
%
\begin{equation}
\label{eq:BH} \mathrm{FDR} = \mathbb{E} \biggl[ \frac{V}{R \vee1} \biggr
] = q
\frac{p_0}{p},
\end{equation}
where $p_0$ is the number of true null hypotheses, $p_0:= \vert\{i:
\beta_i = 0\}\vert= p - \llVert\beta\rrVert_{\ell_0}$.
\end{itemize}
In contrast to Bonferroni's method, BH is an adaptive procedure in the
sense that the threshold for rejection $\vert y\vert_{(\iBH)}$ is
defined in a
data-dependent fashion, and is sensitive to the sparsity and magnitude
of the true signals. In a setting where there are many large
$\beta_j$'s, the last selected variable needs to pass a far less
stringent threshold than it would in a situation where no $\beta_j$ is
truly different from $0$. It has been shown in a variety of papers [see,
e.g., \citet{ABDJ,ABOS,wuzhou,FB}]
that this behavior allows BH to adapt to the unknown
signal sparsity, resulting in some important asymptotic optimality
properties.

We now consider how the Lasso
would behave in this
setting. The solution to
%
%
\begin{equation}
\label{eq:Lasso} \min_{b \in\R^p} \frac{1}{2} \llVert y - Xb
\rrVert^2_{\ell_2} + \lambda\llVert b\rrVert_{\ell_1}
\end{equation}
in the case of orthogonal\vspace*{2pt} designs is given by soft thresholding. In
particular, the Lasso estimate $\hat\beta_j$ is not zero if and only
if $\vert\tilde y_j\vert> \lambda$. That is, variables are
selected using a
nonadaptive threshold $\lambda$. Mindful of the costs associated with
the selection of irrelevant variables, we can control the FWER by
setting $\lambda_{\mathrm{Bonf}}= \sigma\cdot\Phi^{-1}(1-{\alpha
}/{2p}) \approx
\sigma\cdot\sqrt{2 \log p}$.\footnote{For large $t$, we have $1 -
\Phi(t) = t^{-1} \phi(t)(1 + o(t^{-1}))$, where $\phi(\cdot)$
denotes the density of $N(0,1)$. Our approximation comes
from setting the right-hand side to $\alpha/2p$ for a fixed value of
$\alpha$, say, $\alpha= 0.05$, and a large value of $p$.} This
choice, however, is likely to result in a loss of power, and may not
strike the right balance between errors of type I and missed
discoveries. Choosing a value of $\lambda$ substantially smaller than
$\lambda_{\mathrm{Bonf}}$ in a nondata dependent fashion would lead
to a
loss not only of FWER control, but also of FDR control since FDR and FWER
are identical measures under the global null in which all our
variables are irrelevant. Another strategy is to use
cross-validation. However, this data-dependent approach for selecting
the regularization parameter $\lambda$ targets the minimization of
prediction error, and does not offer guarantees with respect to model
selection (see Section~\ref{sec:peak}). Our idea to achieve
adaptivity, thereby increasing power while controlling some form of
type-one error, is to break the monolithic penalty $\lambda
\llVert\beta\rrVert_{\ell_1}$, which treats every variable in the same
manner. Set
\[
\LBH(i) \eqdef\Phi^{-1}(1- q_i), \qquad q_i =
i \cdot q/2p,
\]
and consider the following program:
%
%
\begin{equation}
\label{Def:SlopeBH} \min_{b \in\R^p} \frac{1}{2} \llVert y - Xb
\rrVert^2_{\ell_2} + \sigma\cdot\sum
_{i = 1}^p \LBH(i) \vert b\vert_{(i)},
\end{equation}
where $\vert b\vert_{(1)} \ge\vert b\vert_{(2)} \ge\cdots\ge
\vert b\vert_{(p)}$ are the order
statistics of the absolute values of the coordinates of $b$: in
\eqref{Def:SlopeBH} different variables receive different levels of
penalization depending on their relative importance. While the
similarities of \eqref{Def:SlopeBH} with BH are evident, the solution
to \eqref{Def:SlopeBH} is not a series of scalar-thresholding
operations: the procedures are not---even in this case of orthogonal
variables--exactly equivalent. Nevertheless, an upper bound on FDR
proved in the supplementary appendix [\citet{supple}] can still be assured.
%

\begin{theorem}
\label{teo:fdr_control1}
In the linear model with orthogonal design $X$ and $z \sim
\mathcal{N}(0,\sigma^2 I_n)$, the procedure \eqref{Def:SlopeBH}
rejecting hypotheses for which $\hat\beta_j \neq0$ has an FDR obeying
%
%
\begin{equation}
\label{eq:fdr_control1} \mathrm{FDR} = \mathbb{E} \biggl[\frac
{V}{R \vee1} \biggr] \le q
\frac{p_0}{p}.
\end{equation}
\end{theorem}

\begin{figure}

\includegraphics{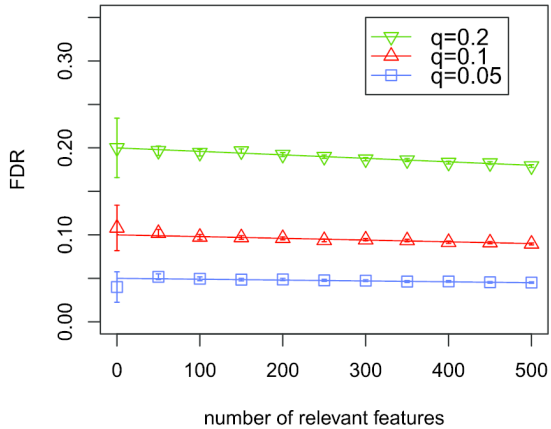}

\caption{FDR of \protect\eqref{Def:SlopeBH} in an orthogonal setting
in which $n=p=5000$. Straight lines correspond to $q \cdot
{p_0}/{p}$, marked points indicate the average False
Discovery Proportion (FDP) across 500 replicates, and bars
correspond to $\pm$2 SE.}\label{Fig_1}
\end{figure}

Figure~\ref{Fig_1} illustrates the FDR achieved by
\eqref{Def:SlopeBH} in simulations using a 5000${}\times{}$5000
orthogonal design $X$ and nonzero regression coefficients equal to
$5\sqrt{2\log p}$.
%
%

We conclude this section with several remarks describing the
properties of our procedure under orthogonal designs:
\begin{longlist}[3.]
\item[1.] While the $\LBH(i)$'s are chosen with reference to BH,
\eqref{Def:SlopeBH} is neither equivalent to the step-up procedure
described above nor to the step-down version.\footnote{The step-down
version rejects $H_{(1)}, \ldots, H_{(i-1)}$, where $i$ is the
first time at which $\vert\tilde y_i\vert/\sigma\le\Phi^{-1}(1-q_i)$.}

\item[2.] The proposal \eqref{Def:SlopeBH} is sandwiched between the
step-down and step-up procedures in the sense that it rejects at
most as many hypotheses as the step-up procedure and at least as
many as the step-down cousin, also known to control the FDR
[\citet{sarkar2002}].

\item[3.] The fact that \eqref{Def:SlopeBH} controls FDR is not a
trivial consequence of this sandwiching.
\end{longlist}
The observations above reinforce the fact that \eqref{Def:SlopeBH} is
different from the procedure known as {\em FDR thresholding} developed
by \citet{AbramovichBenjamini95} in the
context of wavelet estimation and later analyzed in \citet{ABDJ}. With
$t_{\mathrm{FDR}} = \vert\tilde y\vert_{(i_{\mathrm{BH}})}$, FDR
thresholding sets
%
%
\begin{equation}
\label{eq:FDRthresh} \hat\beta_i = \cases{ \tilde y_i, &
\quad$\vert\tilde y_i\vert\ge t_{\mathrm{FDR}}$,
\cr
0, &\quad$\vert
\tilde y_i\vert< t_{\mathrm{FDR}}$.}
\end{equation}
This is a hard-thresholding estimate but with a data-dependent
threshold: the threshold decreases as more components are judged to be
statistically significant. It has been shown that this simple
estimate is asymptotically minimax throughout a range of sparsity
classes [\citet{ABDJ}]. Our method is similar in the sense that it also
chooses an adaptive threshold reflecting the BH procedure. However, it
does not produce a hard-thresholding estimate. Rather, owing to nature
of the sorted $\ell_1$ norm, it outputs a sort of soft-thresholding
estimate. A substantial difference is that FDR thresholding
\eqref{eq:FDRthresh} is designed specifically for orthogonal designs,
whereas the formulation (\ref{Def:SlopeBH}) can be employed for
arbitrary design matrices leading to efficient algorithms. Aside from
algorithmic issues, the choice of the $\lambda$ sequence is, however,
generally challenging.

\subsection{SLOPE}\label{sec1.2}
While orthogonal designs have helped us define the prog\-ram~\eqref{Def:SlopeBH}, this penalized estimation strategy is clearly
applicable in more general settings. To make this explicit, it is
useful to introduce the \textit{sorted $\ell_1$ norm}: letting $\lambda
\neq0$ be a nonincreasing sequence of nonnegative scalars,
%
%
\begin{equation}
\label{eq:lambda} \lambda_1 \ge\lambda_2 \ge\cdots\ge
\lambda_p \ge0,
\end{equation}
we define the sorted-$\ell_1$ norm of a vector $b \in\R^p$
as\footnote{Observe that when all the $\lambda_i$'s take on an
identical positive value, the sorted $\ell_1$ norm reduces to the
usual $\ell_1$ norm (up to a multiplicative factor). Also, when
$\lambda_1 > 0$ and $\lambda_2 = \cdots= \lambda_p = 0$, the sorted
$\ell_1$ norm reduces to the $\ell_\infty$ norm (again, up to a
multiplicative factor).}
%
%
\begin{equation}
\label{eq:orderedl1} J_{\lambda}(b) = \lambda_1 \vert b\vert
_{(1)} + \lambda_2 \vert b\vert_{(2)} + \cdots+
\lambda_p \vert b\vert_{(p)}.
\end{equation}
%
%

\begin{proposition}
\label{prop:orderedl1}
The functional \eqref{eq:orderedl1} is a norm provided
\eqref{eq:lambda} holds.
\end{proposition}

The proof of Proposition~\ref{prop:orderedl1} is provided in the
supplementary appendix
[\citet{supple}]. Now define SLOPE as the solution to
%
%
\begin{equation}
\label{Def:Slope_gen} \mbox{minimize} \qquad\frac{1}{2} \llVert
y-Xb\rrVert
^2 + \sum_{i = 1}^p
\lambda_i \vert b\vert_{(i)}. 
\end{equation}
As a convex program, SLOPE is tractable: as a matter of
fact, we shall see in Section~\ref{sec2} that its computational cost is roughly
the same as that of the Lasso. Just as the sorted $\ell_1$
norm is an extension of the $\ell_1$ norm, SLOPE can be also viewed as
an extension of the Lasso.
SLOPE's general formulation, however,
allows to achieve the adaptivity we discussed earlier.
The case of orthogonal regressors suggests one particular
choice of a $\lambda$ sequence and we will discuss others in later
sections.

\subsection{Relationship to other model selection strategies}\label{sec1.3}

Our purpose is to bring the program \eqref{Def:Slope_gen} to the
attention of the statistical community: this is a computational
tractable proposal for which we provide robust algorithms; it is very
similar to BH when the design is orthogonal, and has promising
properties in terms of FDR control for general designs. We now
compare it with two other commonly used approaches to model selection:
methods based on the minimization of $\ell_0$ penalties and the
adaptive Lasso. We discuss these here because they allow us to
emphasize the motivation and characteristics of the SLOPE algorithm.
We also note that the last few years have witnessed a substantive push
toward the development of an inferential framework after selection
[see, e.g.,
\citet{BY05,POSI,B13,E11}, \citeauthor{JM13J} (\citeyear{JM13J,JM13Jb}), \citet{TibsTaylor,MB10,Meinpvalues,Sara,wasserman,zhangzhang}],
with the exploration
of quite different viewpoints. We will comment on the relationships
between SLOPE and some of these methods, developed while
editing this work, in the discussion section.

\subsubsection{Methods based on \texorpdfstring{$\ell_0$}{ell0} penalties}\label{sec1.3.1}

Canonical model selection procedures find
estimates $\hat\beta$ by solving
%
%
\begin{equation}
\label{eq:l0} \min_{b \in\R^p} \llVert y - Xb\rrVert
_{\ell_2}^2 + \lambda\llVert b\rrVert_{\ell_0},
\end{equation}
where $\llVert b\rrVert_{\ell_0}$ is the number of nonzero
components in $b$. The
idea behind such procedures is to achieve the best possible
trade-off between the goodness of fit and the number of variables
included in the model. Popular selection procedures such as AIC [\citet
{AIC}] and
$C_p$ [\citet{Cp}] are of this form: when the errors are
i.i.d.~$\mathcal{N}(0,\sigma^2)$, AIC and $C_p$ take $\lambda= 2
\sigma^2$. In the high-dimensional regime, such a choice typically
leads to including very many irrelevant variables,
yielding rather poor predictive properties when the
true vector of regression coefficients is sparse. In part to remedy
this problem, \citet{RIC} developed the risk
inflation criterion (RIC): they proposed using
a larger value of $\lambda$, effectively proportional to $2 \sigma^2
\log p $, where $p$ is the total number of variables in
the study. 
Under orthogonal designs, if we associate nonzero fitted coefficients
with rejections, this yields FWER control. Unfortunately, RIC is also
rather conservative and,
therefore, it may not have much power in detecting variables
with nonvanishing regression coefficients unless they are very large.

The above dichotomy has been recognized for some time now and several
researchers have proposed more adaptive strategies. One frequently
discussed idea in the literature is to let the parameter $\lambda$ in
\eqref{eq:l0} decrease as the number of included variables
increases. For instance, when minimizing
\[
\llVert y - X b\rrVert_{\ell_2}^2 + p \bigl(\llVert b\rrVert
_{\ell_0} \bigr),
\]
penalties with appealing information- and decision-theoretic
properties are roughly of the form
%
%
\begin{equation}
\label{eq:pen_adpative} p(k) = 2\sigma^2 k \log(p/k) \quad\mbox
{or}\quad
p(k) = 2\sigma^2 \sum_{1 \le j \le k} \log(p/j).
\end{equation}
Among others, we refer the interested reader to
\citet{FosterStine,BirgeMassart} and to \citet{TibshiraniKnight} for
related approaches.

Interestingly, for large $p$ and small $k$ these penalties are close
to the FDR related penalty
%
%
\begin{equation}
\label{eq:pen_adaptive3} p(k) = \sigma^2 \sum_{1 \le j \le k}
\LBH^2(i),
\end{equation}
proposed in \citet{ABDJ} in the context of the estimation of the vector
of normal means, or regression under the orthogonal design (see the
preceding section) and further explored in \citet{BG09}. Due to an
implicit control of the number of false discoveries, similar model
selection criteria are appealing in gene mapping studies [see, e.g.,
\citet{GWAS2012}]. 

The problem with these selection strategies is that, in
general, they are computationally intractable.
Solving
(\ref{eq:pen_adpative}) would involve a brute-force search essentially
requiring to fit least-squares estimates for {\em all} possible
subsets of variables. This is not practical for even moderate values
of $p$, for example, for $p > 60$.

The decaying
sequence of the smoothing parameters in SLOPE goes along the line of
the adaptive $\ell_0$ penalties specified in \eqref{eq:pen_adpative}, in
which the ``cost per variable included'' decreases as more get
selected. However, SLOPE is computationally tractable and can
be easily evaluated even for large-dimensional problems.


\subsubsection{Adaptive Lasso}\label{sec1.3.2}

Perhaps the most popular alternative to the computationally
intractable $\ell_0$ penalization methods is the Lasso. We have
already discussed some of the limitations of this approach with
respect to FDR control and now wish to explore further the connections
between SLOPE and variants of this procedure. It is well known that
the Lasso estimates of the regression coefficients are biased due to
the shrinkage imposed by the $\ell_1$ penalty. To increase the
accuracy of the estimation of large signals and eliminate some false
discoveries, the adaptive or reweighted versions of Lasso were
introduced [see, e.g., \citet{adlas} or \citet{relas}]. In these
procedures the smoothing parameters $\lambda_1,\ldots,\lambda_p$ are
adjusted to the unknown signal magnitudes based on some estimates of
regression coefficients, perhaps obtained through previous iterations
of Lasso. The idea is then to consider a weighted penalty $\sum_i w_i
\vert b_i\vert$, where $w_i$ is inversely proportional to the estimated
magnitudes so that large regression coefficients are shrunk less than
smaller ones. In some circumstances, such adaptive versions of Lasso
outperform its regular version [\citet{adlas}].

The idea behind SLOPE is entirely different. In the adaptive Lasso,
the penalty tends to decrease as the magnitude of coefficients
increases. In
our approach, the exact opposite happens.
This comes from the fact that
we seek to adapt to the unknown signal sparsity and control FDR. As
shown in \citet{ABDJ}, FDR controlling properties can have interesting
consequences for estimation. In practice, since the SLOPE sequence
$\lambda_1\geq\cdots\geq\lambda_p$ leading to FDR control is
typically rather large, we do not recommend using SLOPE directly for
the estimation of regression coefficients. Instead we propose the
following two-stage procedure: in the first step, SLOPE is used to
identify significant predictors; in the second step, the corresponding
regression coefficients are estimated using the least-squares method
within the identified sparse regression model. Such a two-step
procedure, previously proposed in the context of Lasso [see,
e.g.,~\citet{Meinrelaxed}], can be thought of as an extreme case of
reweighting, where the selected variables are not penalized while
those that are not selected receive an infinite penalty. As shown
below, these estimates have very good properties when the coefficient
sequence $\beta$ is sparse.
\subsubsection{A first illustrative simulation}\label{sec1.3.3}\label{sec:peak}

To concretely illustrate the specific behavior of SLOPE compared to
more traditional penalized approaches, we rely on the simulation of a
relatively simple data structure. We set $n=p=5000$ and generate the
entries of the design matrix with i.i.d.~${\mathcal N}(0,{1}/{n})$
entries. The number of true signals $k$ varies between 0 and 50 and
their magnitudes are set to $\beta_i=\sqrt{2 \log p} \approx4.1$,
while the variance of the error term is assumed known and equal to
1. Since the expected value of the maximum of $p$ independent standard
normal variables is approximately equal to $\sqrt{2\log p}$ and the
whole distribution of the maximum concentrates around this value,
this choice of model parameters makes the sparse signal barely
distinguishable from the noise because the nonzero means are at the
level of the largest null statistics. We refer to, for example,~\citet
{ingster99} for a precise discussion of the limits of
detectability in sparse mixtures.

We fit these observations with three procedures: (1) Lasso with
parameter $\lambda_{\mathrm{Bonf}}=\sigma\cdot\Phi^{-1}(1-{\alpha
}/{2 p})$, which
controls FWER weakly; (2) Lasso with the smoothing parameter $\LCV$
chosen with 10-fold cross-validation; (3) SLOPE with a sequence
$\lambda_1, \ldots, \lambda_p$ defined in Section~\ref{sec:lambda},
expression \eqref{EC}. The level $\alpha$ for $\lambda_{\mathrm
{Bonf}}$ and $q$ for FDR
control in SLOPE are both set to 0.1. To compensate for the fact that
Lasso with $\lambda_{\mathrm{Bonf}}$ and SLOPE tend to apply a much
more stringent
penalization than Lasso with $\LCV$---which aims to minimize
prediction error---we have ``de-biased'' their resulting $\hat{\beta
}$, using ordinary least squares to estimate the
coefficients of the variables selected by Lasso--$\lambda_{\mathrm
{Bonf}}$ and SLOPE [see \citet{Meinrelaxed}].

We compare the procedures on the basis of three criteria: (a) FDR, (b)
power, and (c) relative squared error $\llVert X\hat\beta-
X\beta\rrVert_{\ell^2}^2/\llVert X\beta\rrVert_{\ell_2}^2$.
Note\vspace*{1pt} that only the first
of these measures is meaningful for the case where $k=0$, and in such
a case FDR${}={}$FWER.

%
\begin{figure}[b]

\includegraphics{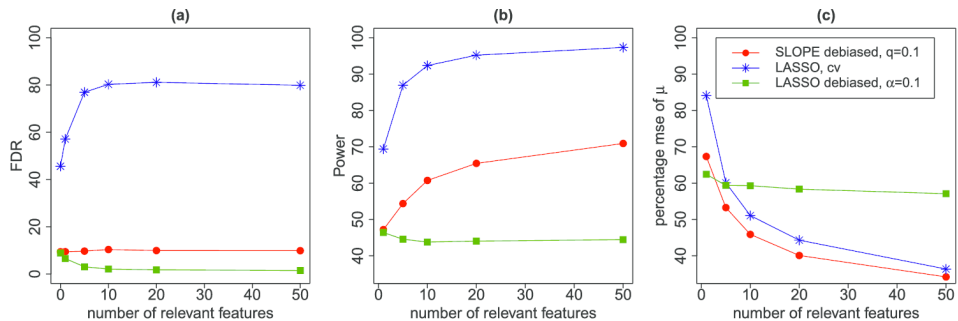}

\caption{Properties of different procedures as a function of the true
number of nonzero regression coefficients: \textup{(a)} FDR, \textup{(b)} power, and \textup{(c)}
relative MSE defined as the average of $100 \cdot\llVert\hat\mu-
\mu\rrVert^2_{\ell_2}/\llVert\mu\rrVert^2_{\ell_2}$, with
$\mu= X\beta$, $\hat\mu
= X\hat\beta$. The design matrix entries are
i.i.d.~$\mathcal{N}(0,{1}/{n})$, $n=p={}$5000, all nonzero regression
coefficients are equal to $\sqrt{2 \log p} \approx4.13$, and
$\sigma^2=1$. Each point in the figures corresponds to the average
of 500 replicates.}\label{fig:com_las}
\end{figure}

Figure~\ref{fig:com_las} reports the results of 500 independent
replicates. The three approaches exhibit quite dramatically different
properties with respect to model selection. SLOPE controls FDR at the
desired level 0.1 for the explored range of $k$; as $k$ increases, its
power goes from 45\% to 70\%. Lasso--$\lambda_{\mathrm{Bonf}}$ has
FDR =0.1 at $k=0$,
and a much lower one for the remaining values of $k$. This results in
a loss of power with respect to SLOPE: irrespective of $k$, the power
is less than 45\%. Cross-validation chooses a $\lambda$ that
minimizes an estimate of prediction error, and in our experiments,
$\LCV$ is quite smaller than a penalization parameter chosen with FDR
control in mind. This results in greater power than SLOPE, but with a
much larger FDR (80\% on average).

%

Figure~\ref{fig:com_las}(c) illustrates the relative mean-square
error, which serves as a measure of prediction accuracy.
It is remarkable how, despite the fact that Lasso--$\LCV$ has higher
power, SLOPE builds a better predictive model since it has a lower
prediction error percentage for all the sparsity levels
considered. 

\section{Algorithms}\label{sec2}
\label{sec:algo}

In this section we present effective algorithms for computing the
solution to SLOPE \eqref{Def:Slope_gen} which rely on the numerical
evaluation of the proximity operator (prox) to the sorted $\ell_1$
norm.

\subsection{Proximal gradient algorithms}\label{sec2.1}
\label{sec:algol}

SLOPE is a convex optimization problem of the form
%
%
\begin{equation}
\label{eq:abstract} \mbox{minimize} \qquad f(b) = g(b) + h(b),
\end{equation}
where $g$ is smooth and convex, and $h$ is convex but not smooth. In
SLOPE, $g$ is the residual sum of squares and, therefore, quadratic,
while $h$ is the sorted $\ell_1$ norm. A~general class of algorithms
for solving problems of this kind are known as {\em proximal gradient
methods}; see \citet{nes07,BoydProx} and references therein. These
are iterative algorithms operating as follows: at each iteration, we
hold a guess $b$ of the solution and compute a local approximation to
the smooth term $g$ of the form
\[
g(b) + \bigl\langle\nabla g(b), x - b \bigr\rangle+ \frac{1}{2t}
\llVert x
- b\rrVert_{\ell_2}^2.
\]
This is interpreted as the sum of a Taylor approximation of $g$ and of
a proximity term; as we shall see, this term is responsible for
searching an update reasonably close to the current guess $b$, and $t$
can be thought of as a step size. Then the next guess $b_+$ is the
unique solution to
\begin{eqnarray*}
b_+ & =& \mathop{\operatorname{arg}\operatorname{min}}_x \biggl\{g(b) + \bigl
\langle\nabla g(b), x - b \bigr\rangle+ \frac{1}{2t} \llVert x -
b\rrVert
_{\ell_2}^2 + h(x) \biggr\}
\\
& =& \mathop{\operatorname{arg}\operatorname{min}}_x \biggl\{
\frac{1}{2t} \bigl\llVert\bigl(b - t \nabla g(b) \bigr) - x \bigr
\rrVert
_{\ell_2}^2 + h(x) \biggr\}
\end{eqnarray*}
(unicity follows from strong convexity).
In the literature, the mapping
\[
x(y) = \mathop{\operatorname{arg}\operatorname{min}}_x \biggl\{
\frac{1}{2t} \llVert y - x\rrVert_{\ell_2}^2 + h(x) \biggr
\}
\]
is called the proximal mapping or prox for short, and denoted by $x =
\operatorname{prox}_{th}(y)$.

The prox of the $\ell_1$ norm is given by entry-wise soft thresholding
[\citet{BoydProx}, page 150] so that a proximal gradient method to solve
the Lasso would take the following form: starting with $b^{0} \in
\R^p$, inductively define
%
\[
b^{k+1} = \eta_{\lambda t_k} \bigl(b^{k} - t_k
X' \bigl(Xb^k - y \bigr); t_k \lambda
\bigr),
\]
where $\eta_{\lambda}(y) = \operatorname{sign}(y) \cdot(\vert
y\vert-
\lambda)_+$
and $\{t_k\}$ is a sequence of step sizes. Hence, we can
solve the Lasso by iterative soft thresholding.

It turns out that one can compute the prox to the sorted $\ell_1$ norm
in nearly the same amount of time as it takes to apply soft
thesholding. In particular, assuming that the entries are sorted (an
order $p\log p$ operation), we
shall demonstrate a linear-time algorithm. Hence, we may consider a
proximal gradient method for SLOPE as in Algorithm~\ref{alg:ista}.

\begin{algorithm}[b]
\caption{Proximal gradient algorithm for SLOPE \protect\eqref{Def:Slope_gen}}
\label{alg:ista}
\begin{algorithmic}[1]
\REQUIRE$b^{0} \in\R^p$
\FOR{$k=0,1,\ldots$}
\STATE$b^{k+1} = \mbox{prox}_{t_k J_\lambda}(b^{k} - t_k X'(Xb^k - y))$
\ENDFOR
\end{algorithmic}
\end{algorithm}

It is well known that the algorithm converges [in the sense that
$f(b^k)$, where $f$ is the objective functional, converges to the
optimal value] under some conditions on the sequence of step sizes
$\{t_k\}$. Valid choices include step sizes obeying $t_k < 2/\llVert
X\rrVert^2$
and step sizes obtained by backtracking line search; see
\citet{tfocs,fista}. Further, one can use duality theory to derive
concrete stopping criteria; see the supplementary Appendix~C [\citet{supple}] for
details.

Many variants are of course possible and one may entertain accelerated
proximal gradient methods in the spirit of FISTA; see \citet{fista} and
\citeauthor{Nesbook} (\citeyear{Nesbook,nes07}). The scheme below is adapted from \citet{fista}.
\begin{algorithm}[t]
\caption{Accelerated proximal gradient algorithm for SLOPE \protect
\eqref{Def:Slope_gen}}
\label{alg:fista}
\begin{algorithmic}[1]
\REQUIRE$b^{0} \in\R^p$, and set $a^{0} = b^{0}$ and $\theta_0=1$
\FOR{$k=0,1,\ldots$}
\STATE$b^{k+1} = \mbox{prox}_{t_k J_\lambda}(a^{k} - t_k X'(Xa^k - y))$
\STATE$\theta_{k+1}^{-1} = \frac{1}2 (1 + \sqrt{1 + 4/\theta_k^2})$
\STATE$a^{k+1} = b^{k+1} + \theta_{k+1}(\theta_k^{-1} - 1)(b^{k+1} - b^k)$
\ENDFOR
\end{algorithmic}
\end{algorithm}

The code in our numerical experiments uses a straightforward
implementation of the standard FISTA algorithm, along with
problem-specific stopping criteria. Standalone Matlab and R
implementations of the algorithm are available at
\surl{http://www-stat.stanford.edu/\textasciitilde candes/SortedL1}. In addition, the
TFOCS package available at \surl{http://cvxr.com} \citet{tfocs}
implements Algorithms~\ref{alg:ista} and~\ref{alg:fista} as well as
its many variants.

\subsection{Fast prox algorithm}\label{sec2.2}
\label{sec:prelim}

Given $y \in\mathbb{R}^p$ and $\lambda_1 \geq\lambda_2 \geq\cdots
\geq\lambda_p \geq0$, the prox to the sorted $\ell_1$ norm is the
unique solution to
%
%
\begin{equation}
\label{eq:prox} \proxh(y; \lambda):= \mathop{\operatorname{arg}\operatorname{min}}_{x \in
\R^p}
\frac{1}{2} \llVert y-x\rrVert_{\ell_2}^2 + \sum
_{i=1}^p \lambda_i \vert x\vert
_{(i)}.
\end{equation}
A simple observation is this: at the solution to \eqref{eq:prox}, the
sign of each $x_i \neq0$ will match that of $y_i$. It therefore
suffices to solve the problem for $\vert y\vert$ and restore the
signs in a
post-processing step, if needed. Likewise, note that applying any
permutation $P$ to $y$ results in a solution $Px$. We can thus choose
a permutation that sorts the entries in $y$ and apply its inverse to
obtain the desired solution. Therefore, without loss of generality, we
can make the following assumption:
%

\begin{assumption}\label{Ass:SortedY}
The vector $y$ obeys $y_1 \geq y_2 \geq\cdots\geq y_p \geq0$.
\end{assumption}

The proposition below, proved in the supplementary Appendix [\citet
{supple}], provides a convenient
reformulation of the proximal problem \eqref{eq:prox} by reformulating
it as a quadratic program (QP).
%

\begin{proposition}\label{alt_formulation}
Under Assumption~\ref{Ass:SortedY} we can reformulate
\eqref{eq:prox} as
%
%
\begin{eqnarray}
\label{Eq:ProxFunSorted}
\mbox{minimize}\qquad&& \frac{1}{2} \llVert y
- x\rrVert
_{\ell_2}^2 + \sum_{i=1}^p
\lambda_i x_i,
\nonumber\\[-8pt]\\[-8pt]\nonumber
\mbox{subject to}\qquad&& x_1 \geq x_2\geq\cdots\geq
x_p \geq0.
\end{eqnarray}
\end{proposition}

We do not suggest performing the prox calculation by calling a
standard QP solver applied to \eqref{Eq:ProxFunSorted}. Rather, we
introduce the FastProxSL1 algorithm for computing the prox: for ease
of exposition, we introduce Algorithm~\ref{alg:fastprox} in its
simplest form before presenting a stack implementation (Algorithm
\ref{Alg:StackAlgorithm}) running in $O(p)$ flops, after an $O(p\log
p)$ sorting step.

\begin{algorithm}[t]
\caption{FastProxSL1}
\label{alg:fastprox}
\begin{algorithmic}
\STATE\textbf{input:} Nonnegative and nonincreasing sequences $y$
and $\lambda$.
\WHILE{$y - \lambda$ is not nonincreasing}
\STATE Identify nondecreasing and nonconstant subsequences, that is, segments $i:j$
such that\vspace*{-2pt}
%
%
\begin{equation}
\label{eq:subseq} y_i - \lambda_i  \leq y_{i+1} - \lambda_{i+1} \leq  \cdots \leq  y_j - \lambda_j\quad\mbox{and}\quad y_i-\lambda_i < y_j - \lambda_j.
\end{equation}
\STATE Replace the values of $y$ and $\lambda$ over such segments
by their average value: for $k \in\{i, i+1, \ldots, j\}$
\[
y_k \gets\frac{1}{j-i+1} \sum_{i \le k \le j}
y_k, \qquad\lambda_k \gets\frac{1}{j-i+1} \sum
_{i \le k \le j} \lambda_k.
\]
\ENDWHILE
\STATE\textbf{output:} $x = (y - \lambda)_+$.
\end{algorithmic}\vspace*{-2pt}
\end{algorithm}

\begin{algorithm}[b]
\caption{Stack-based algorithm for FastProxSL1}
\label{Alg:StackAlgorithm}
\begin{algorithmic}[1]
\STATE\textbf{input:} Nonnegative and nonincreasing sequences $y$
and $\lambda$.
\STATE{\it\# Find optimal group levels}
\STATE$t \gets0$
\FOR{$k = 1$ to $n$}
\STATE$t \gets t + 1$
\STATE$(i,j,s,w)_t = (k, k, y_i - \lambda_i, (y_i - \lambda_i)_+)$
\WHILE{$(t > 1)$ and $(w_{t-1} \leq w_{t})$}
\STATE$(i,j,s,w)_{t-1} \gets(i_{t-1}, j_t, s_{t-1}+s_{t},
(\frac{j_{t-1} - i_{t-1} + 1}{j_{t} - i_{t-1} +1}\cdot s_{t-1}
+ \frac{j_t - i_t + 1}{j_{t} - i_{i-1}+1}\cdot s_{t})_+$)
\STATE Delete $(i,j,s,w)_t$, $t \gets t - 1$
\ENDWHILE
\ENDFOR

\STATE{\it\# Set entries in $x$ for each block}
\FOR{$\ell= 1$ to $t$}
\FOR{$k = i_{\ell}$ to $j_{\ell}$}
\STATE$x_k \gets w_{\ell}$
\ENDFOR
\ENDFOR
\end{algorithmic}\vspace*{-4pt}
\end{algorithm}

Algorithm~\ref{alg:fastprox}, which terminates in at most
$p$ steps, is simple to understand: we simply keep on averaging until
the monotonicity property holds, at which point the solution is known
in closed form. The key point establishing the correctness of the
algorithm is that the update does not change the value of the
prox. This is formalized below.

%
\begin{lemma}
\label{lem:key} The solution does not change after each update;
formally, letting $(y^+, \lambda^+)$ be the updated value of
$(y,\lambda)$ after one pass in Algorithm~\ref{alg:fastprox},
\[
\proxh(y;\lambda) = \proxh\bigl(y^+; \lambda^+ \bigr).
\]
Next, if $(y - \lambda)_+$ is nonincreasing, then it is the solution
to \eqref{eq:prox}, that is, $\proxh(y; \lambda) = (y -\lambda)_+$.
\end{lemma}

This lemma, whose proof is in the supplementary Appendix [\citet
{supple}], guarantees that the
FastProxSL1 algorithm finds the solution to \eqref{eq:prox} in a
finite number of steps.

As stated earlier, it is possible to obtain a careful $O(p)$
implementation of FastProxSL1. Below we present a stack-based
approach. We use tuple notation $(a,b)_i = (c,d)$ to denote $a_i = c$,
$b_i = d$.
For the complexity of the algorithm note that we create a
total of $p$ new tuples. Each of these tuples is merged into a previous
tuple at most once. Since the merge takes a constant amount of time,
the algorithm has the desired ${O}(p)$ complexity.

With this paper, we are making available a C, a Matlab and an R
implementation of the stack-based algorithm at
\url{http://www-stat.stanford.edu/\textasciitilde candes/SortedL1}. The algorithm is
also implemented in R package SLOPE, available on CRAN,
and
included in the current version of the TFOCS
package. Table~\ref{Table:ProxRuntime} reports the average runtimes of
the algorithm (MacBook Pro, 2.66~GHz, Intel Core i7) when applied to
vectors of fixed length and varying sparsity.


%
\begin{table}[b]
\tabcolsep=0pt
\caption{Average runtimes of the stack-based prox implementation
with normalization steps (sorting and sign changes) included,
respectively, excluded}\label{Table:ProxRuntime}
\begin{tabular*}{\tablewidth}{@{\extracolsep{\fill}}@{}lccc@{}}
\hline
& $\bolds{p=10^{5}}$ & $\bolds{p=10^{6}}$ & $\bolds{p=10^{7}}$\\
\hline
Total prox time (s)               & 9.82e--03 & 1.11e--01 & 1.20e$+$00\\
Prox time after normalization (s) & 6.57e--05 & 4.96e--05 & 5.21e--05\\
\hline
\end{tabular*}
\end{table}

\subsection{Related algorithms}\label{sec2.3}

Brad Efron informed us about the connection between the FastProxSL1
algorithm for SLOPE and a simple iterative algorithm for solving
isotonic problems called the pool adjacent violators algorithm (PAVA)
[\citet{kruskal64,barlow72}]. A simple instance of an isotonic regression
problem involves fitting data in a least-squares sense in such a way
that the fitted values are monotone:
%
%
\begin{eqnarray}
\label{eq:isotonic} %
\mbox{minimize} \qquad && \tfrac{1}{2} \llVert y - x\rrVert_{\ell_2}^2,
 \nonumber\\[-8pt]\\[-8pt]\nonumber
\mbox{subject to} \qquad && x_1 \geq x_2\geq\cdots\geq x_p.
\end{eqnarray}
Here, $y$ is a vector of observations and $x$ is the vector of fitted
values, which are here constrained to be nonincreasing. We have chosen
this formulation to emphasize the connection with
\eqref{Eq:ProxFunSorted}. Indeed, our QP \eqref{Eq:ProxFunSorted} is
equivalent to
\begin{eqnarray*}
\mbox{minimize} \qquad && \frac{1}{2} \sum _{i = 1}^p (y_i -
\lambda_i - x_i)^2,
\\
\mbox{subject to} \qquad &&  x_1 \geq x_2\geq\cdots\geq
x_p \ge0,
\end{eqnarray*}
so that we see are really solving an isotonic regression problem with
data \mbox{$y_i - \lambda_i$}. Algorithm~\ref{alg:fastprox} is then a version
of PAVA as described in \citet{barlow72}; see \citet{best90,odersimplex}
for related work and connections with active set methods. Also, an
elegant R package for isotone regression has been contributed by
\citet{deleeuw2009isotone}
and can be used to compute the prox to the sorted $\ell_1$ norm.

Similar algorithms were also proposed in \citet{OSCAR} to solve the
OSCAR optimization problem defined as
%
%
\begin{equation}
\label{eq:OSCAR} \mbox{minimize} \qquad\frac{1}{2} \llVert y-Xb\rrVert
_{\ell_2}^2 + \lambda_1 \llVert b\rrVert
_{\ell_1} + \lambda_2 \sum_{i < j}
\max\bigl(\vert b_i\vert, \vert b_j\vert\bigr).
\end{equation}
The OSCAR formulation was introduced in \citet{OSCAR1} to
encourage grouping of correlated predictors. The OSCAR
penalty term can be expressed as
$\sum_{i=1}^p \alpha_i \vert b\vert_{(i)}$ with
$\alpha_i = \lambda_1 + (p-i) \lambda_2$; hence, this is a
sorted $\ell_1$ norm with a linearly decaying sequence of
weights. \citet{OSCAR1} do not present a special algorithm for
solving \eqref{eq:OSCAR} other than casting the problem as a
QP. In the article \citet{WSL1}, which appeared after our
manuscript was made publicly available, the OSCAR penalty term
was further generalized to a Weigthed Sorted L-one norm, which
coincides with the SLOPE formulation. This latter article
does not discuss statistical properties of this fitting procedure.

\section{Results}\label{sec3}
\label{sec:results}

We now illustrate the performance of our SLOPE proposal in
three different ways. First, we describe a multiple-testing situation
where reducing the problem to a model selection setting and applying
SLOPE assures FDR control, and results in a testing procedure with
appreciable properties. Second, we discuss guiding principles to
choose the sequence of $\lambda_i$'s in general settings, and
illustrate the efficacy of the proposals with simulations. Third, we
apply SLOPE to a data set collected in genetics investigations.

\subsection{An application to multiple testing}\label{sec3.1}
In this section we show how SLOPE can be used as an effective multiple
comparison controlling procedure in a testing problem with a specific
correlation structure. Consider the following situation. Scientists
perform $p = {}$1000 experiments in each of 5 randomly selected
laboratories, resulting in observations that can be modeled as
%
%
\begin{equation}
\label{test_model} y_{i,j} = \mu_i + \tau_j +
z_{i,j}, \qquad1 \le i \le1000, 1 \le j \le5,
\end{equation}
where the laboratory effects $\tau_j$ are i.i.d.~$\mathcal{N}(0,
\sigma^2_\tau)$ random variables and the errors $z_{i,j}$ are
i.i.d. $\mathcal{N}(0,\sigma^2_z)$, with the $\tau$ and $z$ sequences
independent of each other. It is of interest to test whether $H_i:
\mu_i = 0$ versus a two-sided alternative. Averaging the scores over
all five labs results in
\[
\bar{y}_i = \mu_i + \bar{\tau} + \bar{z}_i,
\qquad1 \le i \le1000,
\]
with $\bar{y} \sim\mathcal{N}(\mu,\Sigma)$ and $\Sigma_{i,i} =
\frac{1}{5}(\sigma^2_\tau+\sigma^2_z) = \sigma^2$ and $\Sigma
_{i,j} =
\frac{1}{5} \sigma^2_\tau=\rho$ for $i \neq j$.

The problem has then been reduced to testing if the marginal means of a
multivariate Gaussian vector with equicorrelated entries do not
vanish. One possible approach is to use marginal tests based on $\bar
{y}_i $'s and rely on the Benjamini--Hochberg procedure to control
FDR. That is, we can order $\vert\bar{y}\vert_{(1)} \ge\vert
\bar{y}\vert_{(2)} \ge\cdots
\ge\vert\bar{y}\vert_{(p)}$ and apply the step-up procedure with critical
values equal to $\sigma\cdot\Phi^{-1}(1-iq/2p)$.

Another possible approach is to ``whiten the noise'' and express our
multiple testing problem in the form of a regression equation
%
%
\begin{equation}
\label{test_reg} \tilde y =\Sigma^{-1/2} \bar y = \Sigma^{-1/2}
\mu+ \varepsilon,
\end{equation}
where $\varepsilon\sim\mathcal{N}(0,I_p)$. Treating $\Sigma^{-1/2}$ as
the regression design matrix, our problem is equivalent to classical
model selection: identify the nonzero components of the vector $\mu$
of regression coefficients.\footnote{To be explicit, \eqref{test_reg}
is the basic regression model with $X = \Sigma^{-1/2}$ and $\beta=
\mu$.} Note that while the matrix $\Sigma$ is far from being
diagonal, $\Sigma^{-1/2}$ is diagonally dominant. For example, when
$\sigma^2=1$ and $\rho=0.5$, then $\Sigma^{-1/2}_{i,i}=1.4128$ and
$\Sigma^{-1/2}_{i,j}=-0.0014$ for $i \neq j$. Thus,\vspace*{1pt} every
low-dimensional submodel obtained by selecting few columns of the
design matrix $\Sigma^{-1/2}$ will be very close to orthogonal. In
summary, the transformation~(\ref{test_reg}) reduces the
multiple-testing problem with strongly positively correlated test
statistics to a problem of model selection under approximately
orthogonal design, which is well suited for the application of SLOPE
with the $\lambda_{\mathrm{BH}}$ values.

%
%
\begin{figure}

\includegraphics{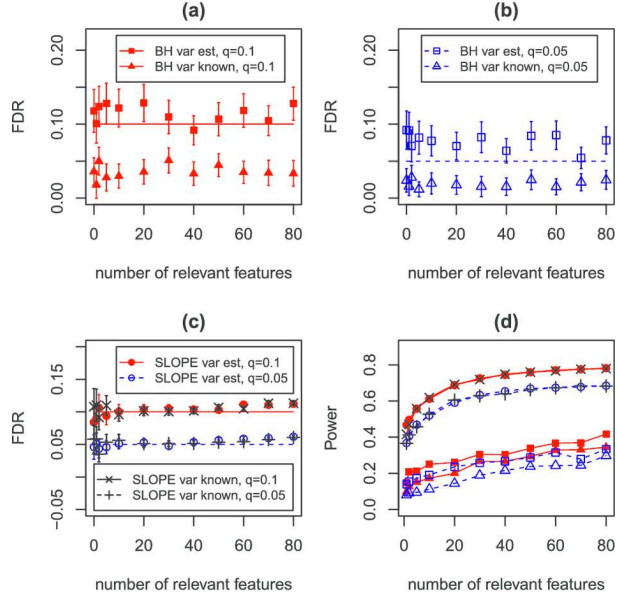}

\caption{Simulation results for testing multiple means from correlated
statistics. \textup{(a)--(b)} Mean FDP $\pm$ 2 SE for marginal tests as a
function of $k$. \textup{(c)} Mean FDP $\pm$ 2 SE for SLOPE. \textup{(d)} Power plot.}\label{fig:anova1}
\end{figure}

To compare the performances of these two approaches, we simulate data
according to the model (\ref{test_model}) with variance components
$\sigma^2_\tau=\sigma^2_z=2.5$, which yield $\sigma^2=1$ and
$\rho=0.5$. We consider a sequence of sparse settings, where the
number $k$ of nonzero $\mu_i$'s varies between $0$ and $80$. To obtain
moderate power, the nonzero means are set to $\sqrt{2\log p}/c\approx
2.63$, where $c$ is\vspace*{1pt} the Euclidean norm of each of the columns of
$\Sigma^{-1/2}$. We compare the performance of SLOPE and BH on
marginal tests under two scenarios: (1) assuming
$\sigma^2_\tau=\sigma^2_z=2.5$ known, and (2) estimating them using
the classical unweighted means method based on equating the ANOVA
mean squares to their expectations:
\[
\hat\sigma_z^2=\mathrm{MSE}, \qquad\hat
\sigma^2_\tau=\frac{\mathrm{MS}\tau-\mathrm{MSE}}{1000};
\]
using the standard notation from ANOVA analysis, MSE is the mean
square due to the error in the model (\ref{test_model}) and MS$\tau$
is the mean square due to the random factor $\tau$.
To use SLOPE, we center the vector $\tilde y$ by subtracting its mean,
and center and standardize the columns of $\hat\Sigma^{-1/2}$, so
they have zero means and unit $l_2$ norms.
Figure~\ref{fig:anova1} reports the results of these simulations,
averaged over 500 independent replicates.



%
\begin{figure}

\includegraphics{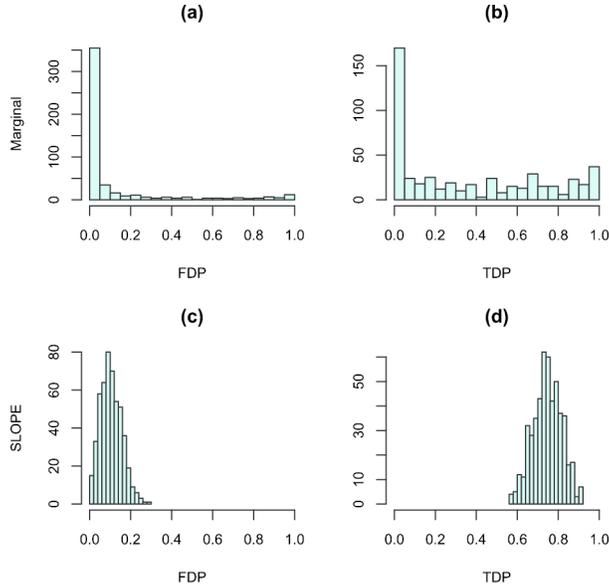}

\caption{Testing example with $q = 0.1$ and $k = 50$. The top row
refers to marginal tests, and the bottom row to SLOPE. Both
procedures use the estimated variance components. Histograms of
false discovery proportions are in the first column and of true
positive proportions in the second.}
\label{fig:anova2}
\end{figure}

In our setting, the estimation procedure has no influence on
SLOPE. Under both scenarios (variance components known and unknown)
SLOPE keeps FDR at the nominal level as long as $k\leq40$. Then its
FDR slowly increases, but for $k\leq80$ it is still very close to the
nominal level as shown in Figure~\ref{fig:anova1}(c). In contrast, the
performance of BH differs significantly: when $\sigma^2$ is known, BH
on the marginal tests is too conservative, with an average FDP below
the nominal level; see Figure~\ref{fig:anova1}(a) and (b). When
$\sigma^2$ is estimated, the average FDP of this procedure increases
and for $q=0.05$, it significantly exceeds the nominal level. Under
both scenarios (known and unknown $\sigma^2$) the power of BH is
substantially smaller than the power provided by SLOPE [Figure~\ref
{fig:anova1}(d)]. Moreover, the False Discovery Proportion (FDP)
in the marginal tests with BH correction appears more variable across
replicates than that of SLOPE [Figure~\ref{fig:anova1}(a),
(b) and~(c)]. Figure~\ref{fig:anova2}
presents the results in greater detail for $q=0.1$ and $k=50$: in
approximately 65\% of the cases the observed FDP for BH is equal to 0,
while in the remaining 35\% it takes values which are distributed
over the whole interval $(0,1)$. This behavior is undesirable. On the
one hand, $\mathrm{FDP} = 0$ typically equates with few discoveries (and
hence power loss). On the other hand, if many $\mathrm{FDP}=0$
contribute to the average in the FDR, this quantity is kept below the
desired level $q$ even if, when there are discoveries, a large number
of them are false. Indeed, in approximately 26\% of all cases BH on
the marginal tests did not make any rejections (i.e., $R=0$); and
conditional on $R>0$, the mean of FDP is equal to 0.16 with a standard
deviation of 0.28, which clearly shows that the observed FDP is
typically far away from the nominal value of $q=0.1$. In other words,
while BH is close to controlling the FDR, the scientists would either
make no discoveries or have very little confidence on those actually
made. In contrast, SLOPE results in a more predictable FDP and a
substantially larger and more predictable True Positive Proportion
(TPP, fraction of correctly identified true signals); see Figure~\ref
{fig:anova2}.

\subsection{Choosing \texorpdfstring{$\lambda$}{lambda} in general settings}\label{sec3.2}
\label{sec:FDR_Lasso}

\begin{figure}

\includegraphics{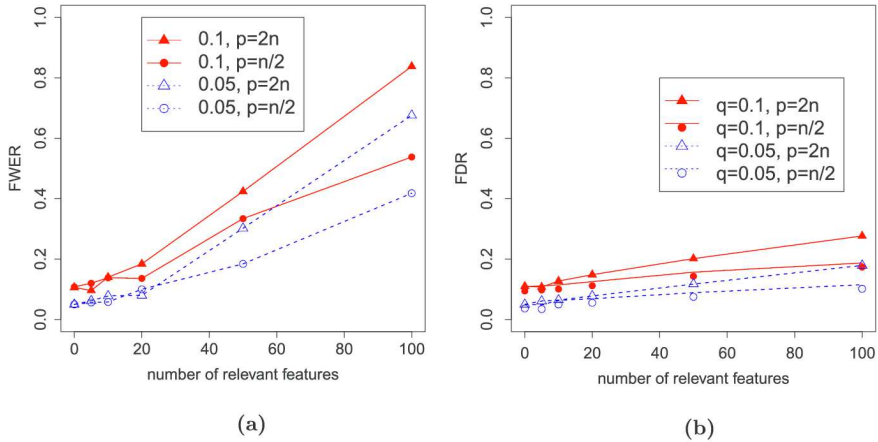}

\caption{Observed \textup{(a)} FWER for Lasso with $\LBONF$  and \textup{(b)} FDR for SLOPE
with $\LBH$  under Gaussian design and $n={}$5000. The results are
averaged over 500 replicates.} \label{fig:Fail}
\end{figure}

In the previous sections we observed that, for orthogonal designs,
Lasso with $\LBONF=\sigma\cdot\Phi^{-1} (1-{\alpha}/{2p} )$
controls FWER at the level $\alpha$, while SLOPE with the sequence
$\lambda=\LBH$ controls FDR at the level $q$. We are interested,
however, in applying these procedures in more general settings,
specifically when $p>n$ and there is some correlation among the
explanatory variables, and when the value of $\sigma^2$ is not
known. We start tackling the first situation. Correlation among
regressors notoriously introduces a series of complications in the
statistical analysis of linear models, ranging from the increased
computational costs that motivated the early popularity of orthogonal
designs, to the conceptual difficulties of distinguishing causal
variables among correlated ones. Indeed, recent results on the
consistency of $\ell_1$ penalization methods typically require some
form of partial orthogonality. SLOPE and Lasso aim at finite sample
properties, but it would not be surprising if departures from
orthogonality were to have a serious effect. To explore this, we
study the performance of Lasso and SLOPE in the case where the entries
of the design matrix are generated independently from the ${\mathcal
N}(0,1/n)$ distribution. Specifically, we consider two Gaussian
designs with $n=5000$: one with $p=2n={}$10,000 and one with
$p=n/2=2500$. We set the value of nonzero coefficients to
$5\sqrt{2\log p}$ and consider situations where the number of
important variables ranges between 0 and 100. Figure~\ref{fig:Fail}
illustrates that under such Gaussian designs
%
both Lasso--$\LBONF$ and SLOPE lose the control over their targeted
error rates (FWER
and FDR)
as the number $k$ of nonzero coefficients increases, with a departure
that is
more severe when the ratio between $p/n$ is larger.
%
%

\subsubsection{The effect of shrinkage}\label{sec3.2.1}
What is behind this fairly strong effect, and is it possible to choose
a $\lambda$ sequence to compensate it? Some useful insights come from
studying the solution of the Lasso.
Assume
that the columns of $X$ have unit norm and that $z\sim{\mathcal
N}(0,1)$. Then the optimality conditions for the Lasso give
%
%
\begin{eqnarray}\label{eq:KKT}
\nonumber
\hat\beta &=& \eta_\lambda\bigl(\hat\beta- X'(X \hat
\beta- y) \bigr) =\eta_\lambda\bigl(\hat\beta- X'(X \hat
\beta- X\beta-z) \bigr)
\nonumber\\[-8pt]\\[-8pt]\nonumber
& =& \eta_\lambda\bigl(\hat\beta- X'X(\hat
\beta-\beta) + X'z \bigr),
\end{eqnarray}
where $\eta_\lambda$ is the soft-thresholding operator,
$\eta_\lambda(t) = \operatorname{sgn}(t) (\vert t\vert- \lambda
)_+$, applied componentwise.
Defining
$ v_i = \langle X_i, \sum_{j \neq i} X_j (\beta_j - \hat\beta
_j)\rangle$, we can write
%
%
\begin{equation}
\hat\beta_i = \eta_\lambda\bigl(\beta_i +
X_i' z + v_i \bigr), \label{eq:KKT2}
\end{equation}
which expresses the relation between the estimated value of
$\hat\beta_i$ and its true value~$\beta_i$. If the variables are
orthogonal, the $v_i$'s are identically equal to $0$, leading to $
\hat\beta_i = \eta_\lambda(\beta_i + X_i' z )$. Conditionally on $X$,
$X_i' z \sim\mathcal{N}(0,1)$ and by using Bonferroni's method, one
can choose $\lambda$ such that $\mathbb{P} (\max_i \vert
X_i'z\vert>\lambda)\le\alpha$.
When $X$ is not orthogonal, however, $v_i \neq0$ and its size
increases with the estimation error of $\beta_j$ (for $i \neq
j$)---which depends on the magnitude of the shrinkage parameter
$\lambda$. Therefore, even in the perfect situation where all the $k$
relevant variables, and those alone, have been selected, and when all
columns of the design matrix are realizations of independent\vspace*{2pt} random
variables, $v_i$ will not be zero. Rather, the squared magnitude $v_i^2$
will be on the order of $\lambda^2 \cdot k/n$.
In other words, the variance that would determine the correct Bonferroni
threshold is on the order $1+ \lambda^2 \cdot k/n$. In reality, the true
$k$ is not known a priori, and the selected $k$ depends on the value
of the smoothing parameter $\lambda$, so that it is not trivial to
implement this correction in the Lasso. SLOPE, however, uses a
decreasing sequence $\lambda$, analogous to a step-down procedure, and
this extra noise due to the shrinkage of relevant variables can be
incorporated by progressively modifying the $\lambda$ sequence. In
evocative, if not exact terms, $\lambda_1$ is used to select the first
variable to enter the model: at this stage we are not aware of any
variable whose shrunk coefficient is ``effectively increasing'' the
noise level, and we can keep $\lambda_1=\LBH(1)$. The value of
$\lambda_2$ determines the second variable to enter the model and,
hence, we know that there is already one important variable whose
coefficient has been shrunk by roughly $\LBH(1)$; we can use this
information to redefine $\lambda_2$. Similarly, when using
$\lambda_3$ to identify the third variable, we know of two relevant
regressors whose coefficients have been shrunk by amounts determined
by $\lambda_1$ and $\lambda_2$, and so on. What follows is an attempt
to make this intuition more precise, accounting for the fact that the
sequence $\lambda$ needs to be determined a priori, and we need to
make a prediction on the values of the cross products $X_i' X_j$
appearing in the definition of $v_i$. Before we turn to this, we want
to underscore how this explanation for the loss
of FDR control is consistent with patterns evident from Figure~\ref
{fig:Fail}: the problem is more serious as $k$ increases (and,
hence, the effect of shrinkage is felt on a larger number of
variables) and as the ratio $p/n$ increases (which for Gaussian
designs results in larger empirical correlation $\vert X_i' X_j\vert
$). Our
loose analysis suggests that when $k$ is really small, SLOPE with
$\LBH$ yields an FDR that is close to the nominal level, as
empirically observed.

\subsubsection{Adjusting the regularizing sequence for SLOPE}\label{sec3.2.2}
\label{sec:lambda}

In light of \eqref{eq:KKT2}, we would like an expression for
$X_i'\XS(\betaS-\hat{\beta}_{\mathcal{S}})$, where with $\mathcal{S}$,
$\XS$ and $\betaS$ we indicate the support of $\beta$, the subset of
variables associated to $\beta_i\neq0$, and the value of their
coefficients, respectively.

Again, to obtain a very rough evaluation of the SLOPE solution, we can
start from the Lasso. Let us assume that the size of $\betaS$ and the
value of $\lambda$ are such that the support and the signs of the
regression coefficients are correctly recovered in the solution. That
is, we assume that $\operatorname{sign}(\beta_j) = \operatorname
{sign}(\hat\beta_j)$
for all $j$, with the convention that $\operatorname{sign}(0) = 0$. Without
loss of generality, we further assume that $\beta_j \ge0$. Now, the
Karush--Kuhn--Tucker (KKT) optimality conditions for the Lasso yield
%
%
\begin{equation}
X'_S(y - X \hat{\beta}_S) = \lambda\cdot
1_S,
\end{equation}
implying
\[
\hat{\beta}_S = \bigl(X'_S X_S
\bigr)^{-1} \bigl(X'_S y - \lambda\cdot
1_S \bigr).
\]
In the case of SLOPE, rather than one $\lambda$, we have a sequence
$\lambda_1,\ldots, \lambda_p$. Assuming again that this is chosen so
that we recover exactly the support $\mathcal{S}$,
the
estimates of the nonzero components are very roughly equal to
\[
\hat{\beta}_S= \bigl(\XS' \XS\bigr)^{-1}
\bigl( \XS' y - \lambdaS\bigr)= \hat\beta_{\mathrm{OLS}} - \bigl(
\XS' \XS\bigr)^{-1}\lambdaS,
\]
where $\lambdaS= (\lambda_1, \ldots, \lambda_{\vert S\vert})'$
and $\hat
\beta_{\mathrm{OLS}}$ is the least-squares estimator of $\beta_S$. This
leads to $\mathbb{E}(\betaS-\hat\beta_S)\approx(\XS' \XS
)^{-1}\lambdaS$ and
\[
\mathbb{E} X_i'\XS(\betaS- \hat{\beta}_{\mathcal{S}})
\approx\mathbb{E} X_i'\XS\bigl(\XS' \XS
\bigr)^{-1} \lambdaS,
\]
an expression that tells us the
typical size of $v_i$ in \eqref{eq:KKT2}.

For the case of Gaussian designs,
where
the entries of $X$ are i.i.d.~$\mathcal{N}(0,1/n)$, for $i \notin
\mathcal{S}$,
%
%
\begin{eqnarray}\label{weights}
\mathbb{E} \bigl(X'_i \XS\bigl(\XS' \XS
\bigr)^{-1} \lambdaS\bigr)^2 &=& \frac{1}{n}
\lambdaS' \mathbb{E} \bigl(\XS' \XS\bigr)^{-1}
\lambdaS= w \bigl(\vert\mathcal{S}\vert\bigr) \cdot\llVert\lambdaS\rrVert
_{\ell_2}^2,
\nonumber\\[-8pt]\\[-8pt]\nonumber
w(k) &=& \frac{1}{n - k - 1}.
\end{eqnarray}
This uses the fact that the expected value of an inverse $k \times k$
Wishart with $n$ degrees of freedom is equal to $I_k/(n-k-1)$.

\begin{figure}

\includegraphics{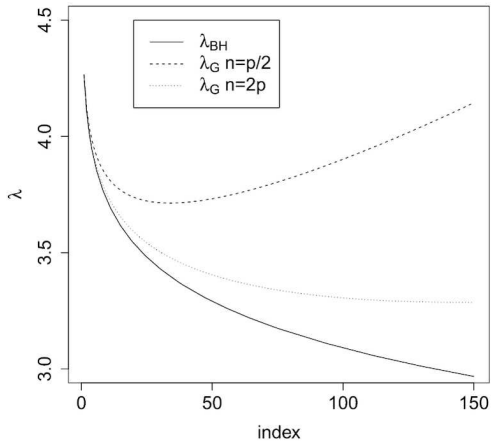}

\caption{Graphical representation of sequences $\{\lambda_i\}$
for $p=5000$ and $q=0.1$. The solid line is $\LBH$, the
dashed (resp., dotted) line is $\LG$ given by
\protect\eqref{eq:correction} for $n=p/2$ (resp., $n =
2p$).} \label{fig:newlambda}
\end{figure}

This suggests the sequence of $\lambda$'s described below denoted by
$\LG$ since it is motivated by Gaussian designs. We start with $\LG(1)
= \LBH(1)$. At the next stage, however, we need to account for the
slight increase in variance so that we do not want to use $\LBH(2)$
but rather
\[
\LG(2) = \LBH(2) \sqrt{1 + w(1) \LG(1)^2}.
\]
Continuing, this gives
%
%
\begin{equation}
\label{eq:correction} \LG(i) = \LBH(i) \sqrt{1 + w(i-1)\sum
_{j < i} \LG(j)^2}.
\end{equation}
Figure~\ref{fig:newlambda} plots the adjusted values given by \eqref
{eq:correction}. As
is clear, these new values yield a procedure that is more conservative
than that based on $\LBH$.
%
%
%
%
%
\begin{figure}

\includegraphics{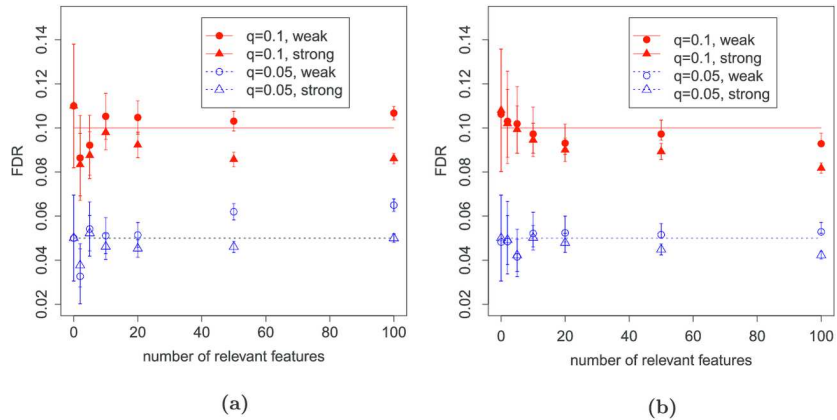}

\caption{Mean FDP $\pm$ 2~SE for SLOPE with $\lambda_{\mathrm
{G}^\star}$. Strong
signals have nonzero regression coefficients set to
$5\sqrt{2\log p}$, while this value is set to $\sqrt{2\log
p}$ for weak signals. \textup{(a)}~$p=2n={}$10,000.
\textup{(b)}~$p=n/2={}$2500.}\label{fig:corrected}
\end{figure}
It can be observed that the corrected sequence $\LG(i)$ may no longer
be decreasing (as in the case where $n = p/2$ in the figure).
It
would not make sense to use such a sequence---note that SLOPE would no
longer be convex---and letting $k^\star= k(n,p,q)$ be the location of
the global minimum, we shall work with
%
%
\begin{equation}
\label{EC} \lambda_{\mathrm{G}^\star}(i) = \cases{ \LG(i), &\quad$i \le
k^\star$,
\cr
\lambda_{k^\star}, &\quad$i > k^\star$,}
\qquad\mbox{with }\LG(i)\mbox{ as in \eqref{eq:correction}}.
\end{equation}
%

An immediate validation---if the intuition that we have stretched this
far has any bearing in reality---is the performance of $ \lambda
_{\mathrm{G}^\star}$ in the
setup of Figure~\ref{fig:Fail}. In Figure~\ref{fig:corrected} we
illustrate the performance of SLOPE for large signals
$\beta_i=5\sqrt{2 \log p}$ as in Figure~\ref{fig:Fail}, as well as for
rather weak signals with $\beta_i=\sqrt{2\log p}$. The correction
works very well, rectifying the loss of FDR control documented in
Figure~\ref{fig:Fail}. For $p= 2n ={}$10,000, the values of the
critical point $k^\star$ are $51$ for $q = 0.05$ and $68$ for $q =
0.1$. For $p=n/2={}$2500, they become $95$ and $147$, respectively. It
can be observed that for large signals, SLOPE keeps FDR below the
nominal level even after passing the critical point. Interestingly,
the control of FDR is more difficult when the coefficients have small
amplitudes. We believe that some increase of FDR for weak signals is
related to the loss of power, which our correction does not account
for. However, even for weak signals the observed FDR of SLOPE with
$\lambda_{\mathrm{G}^\star}$ is very close to the nominal level when
$k\leq k^\star$.

%
%

In situations where one cannot assume that the design is Gaussian or
that columns are independent,\vspace*{1pt} we suggest replacing $w(i-1)\sum_{j < i}
\lambda_j^2$ in the formula~(\ref{eq:correction}) with a Monte Carlo
estimate of the correction. Let $X$ denote the standardized version
of the design matrix, so that each column has a mean equal to zero and
unit $l_2$ norm. Suppose we have computed $\lambda_1, \ldots,
\lambda_{i-1}$ and wish to compute $\lambda_i$. Let $X_{\mathcal{S}}$
indicate a matrix formed by selecting those columns with indices in
some set $\mathcal{S}$ of cardinality $i-1$ and let $j \notin
\mathcal{S}$. After\vspace*{1pt} randomly selecting $\mathcal{S}$ and $j$, the
correction can be approximated by the average of $(X_j'
X_{\mathcal{S}} (X_{\mathcal{S}}' X_{\mathcal{S}})^{-1}
\lambda_{1:i-1})^2$ across realizations, where
$\lambda_{1:i-1}=(\lambda_1,\ldots,\lambda_{i-1})'$.

Significantly more research is needed to understand the properties of
this heuristic and to design more efficient alternatives. Our
simulations so far suggest that it provides approximate FDR control
when looking at the average across all possible signal placements,
and---for any fixed signal location---if the columns of the design
matrix are exchangeable. It is important to note that the
computational cost of this procedure is relatively low. Two elements
contribute to this. First, the complexity of the procedure is reduced
by the fact that the sequence of $\lambda$'s does not need to be
estimated entirely, but only up to the point $k^{\star}$ where it
starts increasing (or simply flattens) and only for a number of entries
on the order of the expected number of nonzero coefficients. Second,
the smoothness of $\lambda$ assures that it is enough to estimate
$\lambda$ on a grid of points between 1 and $k^{\star}$, making the
problem tractable also for very large $p$. In \citet{SLOPE} we applied
a similar procedure for the estimation of the regularizing sequence
with $p=2048^2={}$4,194,304 and $n=p/5$ and found out that it was
sufficient to estimate this sequence at only 40 grid points.

\subsubsection{Unknown \texorpdfstring{$\sigma$}{sigma}}\label{sec3.2.3}

According to formulas (\ref{Def:SlopeBH}) and (\ref{Def:Slope_gen}),
the penalty in SLOPE depends on the standard deviation $\sigma$ of the
error term. In many applications $\sigma$ is not known and needs to
be estimated. When $n$ is larger than $p$, this can easily be done by
means of classical unbiased estimators. When $p\geq n$, some solutions
for simultaneous estimation of $\sigma$ and regression coefficients
using $\ell_1$ optimization schemes were proposed; see, for
example,~\citet{ScaledLasso} and \citet{sun2012scaled}. Specifically,
\citet{sun2012scaled} introduced a simple iterative version of the
Lasso called the {\em scaled Lasso}. The idea of this algorithm can be
applied to SLOPE, with some modifications. For one, our simulation
results show that, under very sparse scenarios, it is better to
de-bias the estimates of regression parameters by using classical
least-squares estimates within the selected model to obtain an
estimate of $\sigma^2$.

We present our algorithm above (Algorithm~\ref{alg:sigma}). There, $\lambda^S$ is the sequence of
SLOPE parameters designed to work with $\sigma=1$, obtained using the
methods from Section~\ref{sec:lambda}.
\begin{algorithm}[t]
\caption{Iterative SLOPE fitting when $\sigma$ is unknown}
\label{alg:sigma}
\begin{algorithmic}[1]
\STATE\textbf{input:} $y$, $X$ and initial sequence $\lambda^S$
(computed for $\sigma= 1$)
%

%

\STATE\textbf{initialize:} $S_+=\varnothing$
\REPEAT

\STATE$S=S_+$
\STATE compute the RSS obtained by regressing $y$ onto variables in
$S$
\STATE set $\hat{\sigma}^2 = \mathrm{RSS}/(n-\vert S\vert-1)$\vspace*{1pt}
\STATE compute the solution $\hat\beta$ to SLOPE with parameter
sequence $\hat\sigma\cdot\lambda^S$

\STATE set $S_+ = \operatorname{supp}(\hat\beta) $
\UNTIL{$S_+ = S$}
\end{algorithmic}
\end{algorithm}

The procedure starts by using a conservative estimate of the
standard deviation of the error term $\hat\sigma^{(0)} =
\operatorname{Std}(y)$ and a related conservative version of SLOPE
with $\lambda^{(0)} =\hat\sigma^{(0)} \cdot\lambda^S$. Then, in
consecutive runs $\hat\sigma^{(k)}$ is computed using residuals from
the regression model, which includes variables identified by SLOPE
with sequence $\sigma^{(k-1)} \cdot\lambda^S$. The procedure is
repeated until convergence, that is, until the next iteration results in
exactly the same model as the current one.

\subsubsection{Simulations with idealized GWAS data}\label{sec3.2.4}


We illustrate the performance of the ``scaled'' version of SLOPE
and of our algorithm
for the estimation of the parameters $\lambda_i$ with simulations
designed to mimic an idealized version of Genome Wide Association
Studies (GWAS).
We set $n=p=5000$, and simulate 5000 genotypes of $p$ independent
Single Nucleotide Polymorphisms (SNPs). For each of these SNPs the
minor allele frequency (MAF) is sampled from the uniform
distribution on the interval $(0.1,0.5)$. Let us underscore that
this assumption of independence is not met in actual GWAS, where the
number of typed SNPs is in the order of millions. Rather, one can
consider our data-generating mechanism as an approximation of the
result of preliminary screening of genotype variants to avoid
complications due to correlation. Our goal here is not to argue that
SLOPE has superior performance in GWAS, but rather to illustrate the
computational costs and inferential results of our algorithms.
The explanatory variables are defined as
%
%
\begin{equation}
\tilde x_{ij} = \cases{ -1, &\quad for $aa$,
\cr
0, &\quad for $aA$,
\cr
1, &\quad for $AA$,}
\end{equation}
where $a$ and $A$ denote the minor and reference alleles at the
$j$th SNP for the $i$th individual. Then the matrix $\tilde X$ is
centered and standardized, so the columns of the final design
matrix $X$ have zero mean and unit norm. The trait values are
simulated according to the model
%
%
\begin{equation}
y=X\beta+z, \label{lmgwas}
\end{equation}
where
$z \sim N(0, I)$, that is, we assume only additive effects and no
interaction between loci (epistasis). We vary the number of
nonzero regression coefficients $k$ between 0 and 50 and we set
their size to $1.2 \sqrt{2\log p} \approx4.95$ (``moderate''
signal). For each value of $k$, 500 replicates are performed,
in each selecting randomly among the columns of $X$, the $k$ with
nonzero coefficients. Since our design matrix is centered and
does not contain an intercept, we also center the vector of
responses and let SLOPE work with $\tilde y=y-\bar y$, where $\bar
y$ is the mean of $y$.

\begin{figure}

\includegraphics{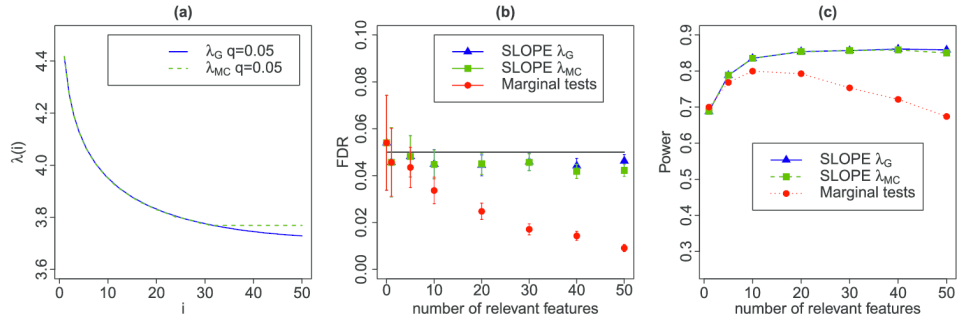}

\caption{\textup{(a)} Graphical representation of sequences
$\LMC$ and $\lambda_{\mathrm{G}}$ for the SNP design matrix. \textup{(b)}~Mean
FDP $\pm$ 2~SE for SLOPE with $\lambda_{\mathrm{G}^\star}$ and
$\LMC$ and for BH as applied to marginal tests. \textup{(c)}~Power of both
versions of SLOPE and BH on marginal tests for
$\beta_1=\cdots=\beta_k=1.2 \sqrt{2\log p}\approx4.95$, $\sigma
=1$. In each replicate, the
signals are randomly placed over the columns of the design
matrix, and the plotted data points are averages over 500
replicates.}
\label{Figure8_revised}
\end{figure}

We set $q = 0.05$ and estimate the sequence $\lambda$ via the
Monte Carlo approach described in Section~\ref{sec:lambda}; here,
we use 5000 independent random draws of $X_{\mathcal{S}}$ and
$X_j$ to compute the next term in the sequence. The calculations
terminated in about 90 seconds (HP EliteDesk 800 G1 TWR, 3.40~GHz,
Intel i7-4770) at $\lambda_{31}$, where the estimated sequence
$\lambda$ obtained a first local minimum. Figure~\ref
{Figure8_revised}(a) illustrates that up to this first minimum the
Monte Carlo sequence
$\LMC$ coincides with the heuristic sequence
$\lambda_{\mathrm{G}^\star}$ for Gaussian matrices. In the result
the FDR and power of ``scaled'' SLOPE are almost the same
for both sequences [Figure~\ref{Figure8_revised}(b) and~(c)].

%

In our simulations, the proposed algorithm for scaled SLOPE
converges very quickly. The conservative initial estimate of
$\sigma$ leads to a relatively small model with few false
discoveries since $\sigma^{(0)}\cdot\lambda^S$ controls the FDR in
sparse settings. Typically, iterations to convergence see the
estimated value of $\sigma$ decrease and the number of selected
variables increase. Since some signals remain undetected (the power
is usually below 100\%), $\sigma$ is slightly overestimated at the
point of convergence, which translates into controlling the FDR at a
level slightly below the nominal one; see Figure~\ref{Figure8_revised}(b).


Figure~\ref{Figure8_revised}(b) and (c) compare scaled SLOPE with
the ``marginal'' tests. The latter are based on $t$-test statistics
\[
t_i=\hat\beta_i/\hat\sigma^2, \qquad\hat
\sigma^2=\mathrm{RSS}_i/(n-2),
\]
where $\hat\beta_i$ (resp., $\mathrm{RSS}_i$) is the least-square
estimate of
the regression coefficient (resp., the residual sum of squares) in the
simple linear regression model including only the $i$th SNP. To
adjust for multiplicity, we use BH at the nominal FDR level $q=0.05$.

It can be observed
that SLOPE and marginal tests do not differ substantially when $k\leq
5$. However, for $k\geq10$ the FDR of the marginal tests approach
falls below the nominal level and the power decreases from 80\% for
$k=10$ to 67\% for $k=50$. SLOPE's power remains, instead, stable
at the level of approximately 86\% for $k\in\{20,\ldots,50\}$. This
conservative behavior of marginal tests results from the inflation of
the noise level
estimate caused by regressors that are unaccounted for in the simple
regression model.

We use this idealized GWAS setting to also explore the effect
of some model misspecification. First, we consider a trait $y$
on which genotypes have effects that are not simply additive. We
formalize this via the matrix $\tilde Z$ collecting the ``dominant''
effects
%
%
\begin{equation}
\tilde z_{ij} = \cases{ -1, &\quad for $aa,AA$,
\cr
1, &\quad for
$aA$.}
\end{equation}
The final design matrix $[X, Z]$ has the columns $[\tilde X, \tilde
Z]$ centered and standardized. Now the trait values are simulated
according to the model
\[
y=[X,Z] \bigl[\beta'_X,\beta'_Z
\bigr]'+\varepsilon,
\]
where $\varepsilon\sim N(0, I)$, the number of ``causal'' SNPs $k$ varies
between 0 and 50, each causal SNP has an additive effect (nonzero
components of $\beta_X$) equal to $1.2 \sqrt{2\log p} \approx4.95$
and a dominant effect (nonzero components of $\beta_Z$) randomly
sampled from $N(0, \sigma=2\sqrt{2 \log p})$. The data is analyzed
using model (\ref{lmgwas}), that is, assuming linear effect of alleles
even when this is not true.

Second, to explore the sensitivity to violations of the assumption
of the normality of the error terms, we considered (1) error terms
$z_i$ with a Laplace distribution and a scale parameter adjusted to
that the variance is equal to one, and (2) error terms contaminated
with 50 outliers $\sim N(0,\sigma=5)$ representing 1\% of all
observations.

%
\begin{figure}

\includegraphics{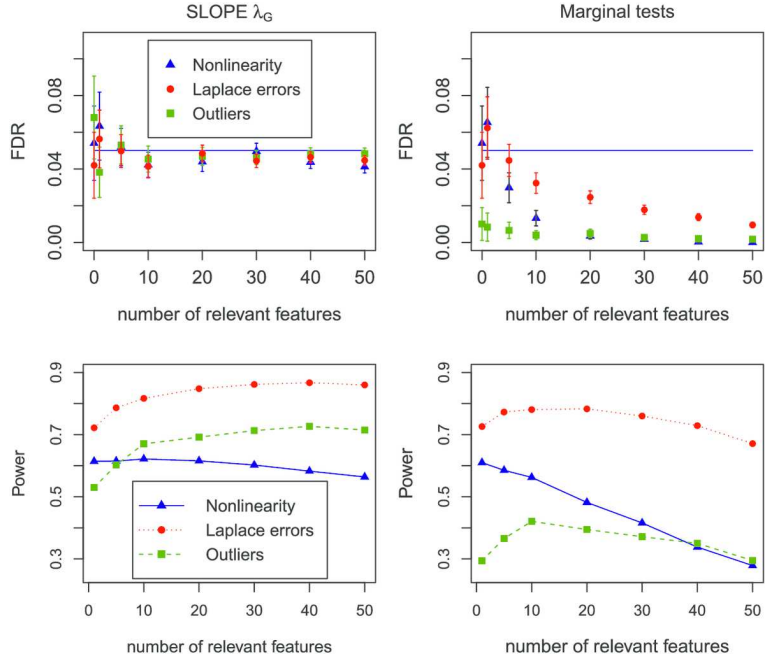}

\caption{FDR and power of ``scaled'' SLOPE based on ``gaussian''
sequence $\lambda_{\mathrm{G}^\star}$ (left panel) and BH-corrected
single marker tests (right panel) for different deviations from the
assumed regression model. Error bars for FDR correspond to mean FDP
$\pm$ 2 SE.}
\label{Figure9_revised}
\end{figure}

Figure~\ref{Figure9_revised} summarizes the performance of SLOPE and
of the marginal tests (adjusted for multiplicity via BH), which we
include for reference purposes. Violation of model assumption appears
to affect power rather than FDR in the case of SLOPE. Specifically, in
all three examples FDR is kept very close to the nominal level while
the power is somewhat diminished with respect to Figure~\ref
{Figure8_revised}. The smallest
difference is observed in the case of Laplace errors, where the results
of SLOPE are almost the same as in the case of normal errors. This is
also the case where the difference in performance due to model
misspecification is negligible for marginal tests. In all other
cases, this approach seems to be much more sensitive than SLOPE to
model misspecification.

\subsection{A real data example from genetics}\label{sec3.3}\label{Gen}

In this section we illustrate the application of SLOPE to a current
problem in genetics. In \citet{SetF14}, the authors investigate the
role of genetic variants in 17 regions in the genome, selected on the
basis of previously reported association with traits related to
cardiovascular health. Polymorphisms are identified via exome
resequencing in approximately 6000 individuals of Finnish descent:
this provides a comprehensive survey of the genetic diversity in the
coding portions of these regions and affords the opportunity to
investigate which of these variants have an effect on the traits of
interest. While the original study has a broader scope, we here tackle
the problem of identifying which genetic variants in these regions
impact the fasting blood HDL levels. Previous literature reported
associations between 9 of the 17 regions and HDL, but the resolution
of these earlier studies was unable to pinpoint to specific variants
in these regions or to distinguish if only one or multiple variants
within the regions impact HDL. The resequencing study was designed to
address this problem.

The analysis in \citet{SetF14} relies substantially on ``marginal''
tests: the effect of each variant on HDL is examined via a linear
regression that has cholesterol level as outcome and the genotype of
the variant as explanatory variable, together with covariates that
capture possible population stratification. Such marginal tests are
common in genetics and represent the standard approach in genome-wide
association studies (GWAS). Among their advantages, it is worth
mentioning that they allow to use all available observations for each
variant without requiring imputation of missing data; their
computational cost is minimal; and they result in a $p$-value for each
variant that can be used to clearly communicate to the scientific
community the strength of the evidence in favor of its impact on a
particular trait. Marginal tests, however, cannot distinguish if the
association between a variant and a phenotype is ``direct'' or due to
correlation between the variant in question and another, truly linked
to the phenotype. Since most of the correlation between genetic
variants is due to their location along the genome (with nearby
variants often correlated), this confounding is often considered not
too serious a limitation in GWAS: multiple polymorphisms associated to
a phenotype in one locus simply indicate that there is at least one
genetic variant (most likely not measured in the study) with impact on
the phenotype in the locus. The situation is quite different in the
resequencing study we want to analyze, where establishing if one or
more variants in the same region influence HDL is one of the goals. To
address this, the authors of \citet{SetF14} resort to regressions that
include two variables at the time: one of these being the variant with
previously documented strongest marginal signal in the region, the
other being variants that passed an FDR controlling threshold in the single
variant analysis. Model selection strategies were only cursorily
explored with a step-wise search routine that targets BIC. Such
limited foray into model selection is motivated by the fact that one
major concern in genetics is to control some global measure of type I
error, and currently available model selection strategies do not offer
finite sample guarantees with this regard. This goal is in line with
that of SLOPE and so it is interesting for us to apply this new
procedure to this problem.

The data set in \citet{SetF14} comprises 1878 variants, on 6121
subjects. Before analyzing it with SLOPE, or other model selection
tools, we performed the following filtering. We eliminated from
considerations variants observed only once (a total of 486), since it
would not be possible to make inference on their effect without strong
assumptions. We examined correlation between variants and selected for
analysis a set of variants with pair-wise correlation smaller than
0.3. Larger values would make it quite challenging to interpret the
outcomes; they render difficult the comparison of results across
procedures since these might select different variables from a group
of correlated ones; and large correlations are likely to adversely
impact the efficacy of any model selection procedure. This reduction
was carried out in an iterative fashion, selecting representatives from
groups of correlated variables, starting from stronger levels of
correlation and moving onto lower ones. Among correlated variables, we
selected those that had stronger univariate association with HDL,
larger minor allele frequency (diversity), and, among very rare
variants, we privileged those whose annotation was more indicative of
possible functional effects. Once variables were identified, we
eliminated subjects that were missing values for more than 10
variants and for HDL. The remaining missing values were imputed using
the average allele count per variant. This resulted in a design with
5375 subjects and 777 variants. The minor allele frequency of the
variants included ranges from $2\times10^{-4}$ to 0.5, with a median
of 0.001 and a mean of 0.028: the data set still includes a number of
rare variants, with the minor allele frequency smaller than 0.01.

\begin{figure}

\includegraphics{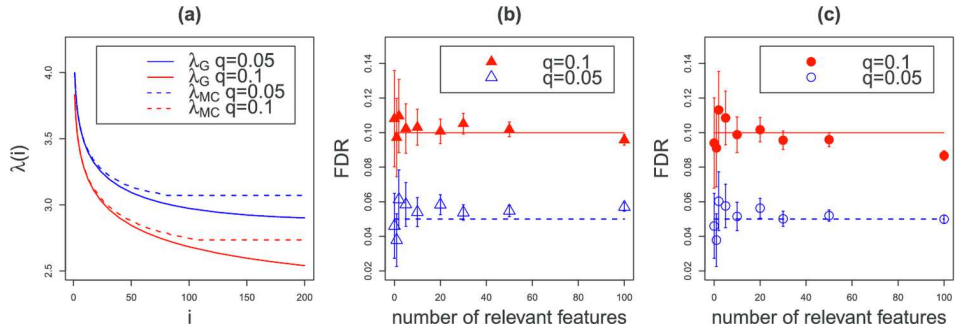}

\caption{\textup{(a)} Graphical representation of sequences
$\LMC$ and $\lambda_{\mathrm{G}}$ for the variants design matrix.
Mean FDP $\pm$ 2~SE for SLOPE with \textup{(b)} $\lambda_{\mathrm{G}^\star}$
and \textup{(c)} $\LMC$ for the variants design matrix and
$\beta_1=\cdots=\beta_k=\sqrt{2\log p}\approx3.65$, $\sigma=1$.}\vspace*{-5pt}\label{fig:fdr_Chiara}
\end{figure}

In \citet{SetF14}, association between HDL and polymorphisms was
analyzed only for variants in regions previously identified as having
an influence on HDL: \textit{ABCA1}, \textit{APOA1}, \textit{CEPT}, \textit{FADS1},
\textit{GALNT2}, \textit{LIPC}, \textit{LPL}, \textit{MADD}, and
\textit{MVK} (regions are identified with the name of
one of the genes they contain). Moreover, only variants with minor
allele frequencies larger than 0.01 were individually investigated,
while nonsynonimous rare variants were analyzed with ``burden
tests.'' These restrictions were motivated, at least in part, by the
desire to reduce tests to the most well-powered ones, so that
controlling for multiple comparisons would not translate in an
excessive decrease of power. Our analysis is based on all variants
that survive the described filtering in all regions, including those
not directly sequenced in the experiment in \citet{SetF14}, but
included in the study as landmarks of previously documented
associations ({\em array SNPs} in the terminology of the paper). We
compare the following approaches: the (1) marginal tests described
above in conjunction with BH and $q=0.05$; (2) BH and $q=0.05$ applied
to the $p$-values from the full model regression; (3) Lasso with
$\lambda_{\mathrm{Bonf}}$ and $\alpha=0.05$; (4) Lasso with $\LCV$
(in these last two
cases we use the routines implemented in {\tt glmnet} in R); (5) the R
routine Step.AIC in forward direction and BIC as optimality criteria;
(6) the R routine Step.AIC in backward direction and BIC as
optimality criteria; (7) SLOPE with $\lambda_{\mathrm{G}^\star}$ and
$q=0.05$; (8) SLOPE
with $\lambda$ obtained via Monte Carlo starting from our design
matrix. Defining the $\lambda$ for Lasso--$\lambda_{\mathrm{Bonf}}$
and SLOPE requires
a knowledge of the noise level $\sigma^2$; we estimated this from the
residuals of the full model. When estimating $\lambda$ via the Monte
Carlo approach, for each $i$ we used 5000 independent random draws of
$X_{\mathcal{S}}$ and $X_j$. Figure~\ref{fig:fdr_Chiara}(a) illustrates
that the Monte Carlo sequence $\LMC$ is only slightly larger than
$\lambda_{\mathrm{G}^\star}$: the
difference increases with the index $i$, and becomes substantial for
ranges of $i$ that are unlikely to be relevant in the scientific
problem at hand.

%
%

\begin{figure}

\includegraphics{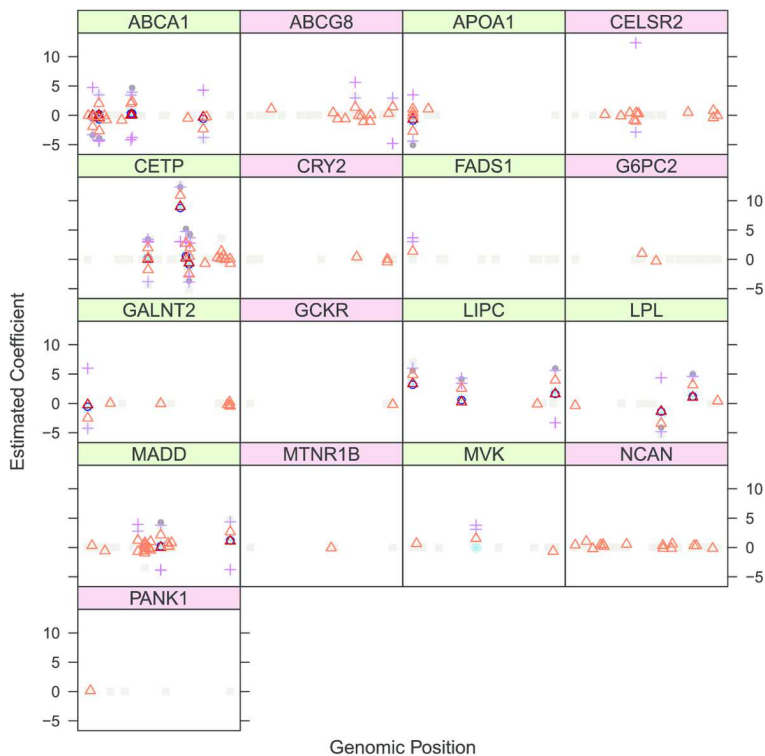}

\caption{Estimated effects on HDL for variants in 17 regions. Each
panel corresponds to a region and is identified by the name of a gene
in the region, following the convention in \citet{SetF14}. Regions with
(without) previously reported association to HDL are on the green (red)
background. On the $x$-axis variants position in base-pairs along their
respective chromosomes. On the $y$-axis estimated effect according to
different methodologies. With the exception of marginal tests---which
we use to convey information on the number of variables and indicated
with light gray squares---we report only the value of nonzero
coefficients. The rest of the plotting symbols and color convention is
as follows: dark gray bullet---BH on $p$-values from full model; magenta
cross---forward BIC; purple cross---backward BIC; red
triangle---Lasso--$\lambda_{\mathrm{Bonf}}$; orange
triangle---Lasso--$\LCV$; cyan star---SLOPE--$\lambda_{\mathrm
{G}^\star}$; black circle---SLOPE with $\lambda$ defined with Monte
Carlo strategy.} \label{fig:genetics1}
\end{figure}

Tables~1 and 2 in \citet{SetF14} describe a
total of 14 variants as having an effect on HDL: two of these are for
regions {\em FADS1} and {\em MVK} and the strength of the evidence in this
specific data set is quite weak (a marginal $p$-value of the order of
$10^{-3}$). Multiple effects are identified in regions \textit{ABCA1}, \textit{CEPT},
\textit{LPL} and \textit{LIPL}. The results of the various ``model selection''
strategies we explored are in Figure~\ref{fig:genetics1}, which
reports the estimated values of the coefficients. The effect of the
shrinkage induced by Lasso and SLOPE are evident. To properly compare
effect sizes across methods, it would be useful to resort to the
two-step procedure that we used for the simulation described in
Figure~\ref{fig:com_las}. Since our interest here is purely model selection,
we report the coefficients directly as estimated by the $\ell_1$
penalized procedures; this has the welcome side effect of increasing
the spread of points in Figure~\ref{fig:genetics1}, improving
visibility.

%

Of the 14 variants described in \citet{SetF14}, 8 are selected by all
methods. The remaining 6 are all selected by at least some of the 8
methods we compared. There are an additional 5 variants that are selected
by all methods but are not in the main list of findings in the original
paper: four of these are rare variants, and one is an {\em array SNP}
for a trait other than HDL. While none of these, therefore, was
singularly analyzed for association in \citet{SetF14}, they are in
highlighted regions: one is in {\em MADD}, and the others in {\em
ABCA1} and {\em CETP},
where the paper documents a plurality of signals.

Besides this core of common selections that correspond well to the
original findings, there are notable differences among the 8
approaches we considered. The total number of selected variables
ranges from 15, with BH on the $p$-values of the full model, to 119,
with the cross-validated Lasso. It is not surprising that these
methods would result in the extreme solutions. On the one hand, the
$p$-values from the full model reflect the contribution of one variable
given all the others, which are, however, not necessarily included in
the models selected by other approaches; on the other hand, we have
seen how the cross-validated Lasso tends to select a much larger
number of variables and offers no control of FDR. In our case, the
cross-validated Lasso estimates nonzero coefficients for 90 variables
that are not selected by any other methods. Note that the number of
variables selected by the cross-validated Lasso changes in different
runs of the procedure, as implemented in {\tt glmnet} with default
parameters. It is quite reasonable to assume that a large number of
these are false positives: regions \textit{G6PC2}, \textit{PANK1}, \textit{CRY2} and \textit{MTNR1B},\vadjust{\goodbreak} where
the Lasso--$\LCV$ selects some variants, have no documented association
with lipid levels, and regions \textit{CELSR2}, \textit{GCKR}, \textit{ABCG8} and \textit{NCAN} have been
associated previously to total cholesterol and LDL, but not HDL. The
other procedures that select some variants in any of these regions are
the forward and backward greedy searches trying to optimize BIC, which
have hits in \textit{CELSR2} and \textit{ABCG8}, and the BH on univariate
$p$-value, which
has one hit in \textit{ABCG8}. SLOPE does not select any variant in
regions not
known to be associated with HDL. This is true also of the
Lasso--$\lambda_{\mathrm{Bonf}}$ and BH on the $p$-values from the full
model, but these
miss, respectively, 2 and 6 of the variants described in the original
paper, while SLOPE $\lambda_{\mathrm{G}^\star}$ misses only one of them.

%
\begin{figure}

\includegraphics{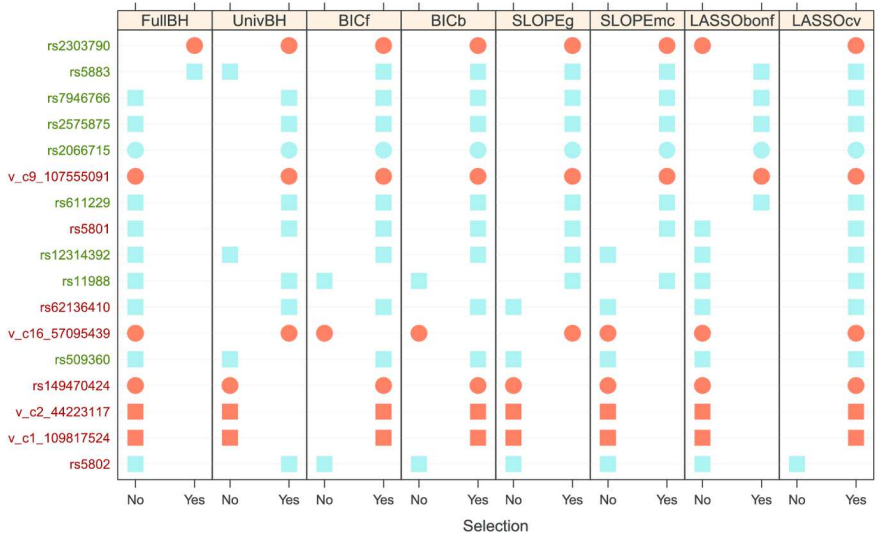}

\caption{Each row corresponds to a variant in the set differently
selected by the compared procedures, indicated by columns. Orange
is used to represent rare variants and blue common ones. Squares
indicate synonymous (or noncoding variants) and circles nonsynonimous
ones. Variants are ordered according to the frequency with which they
are selected. Variants with names in green are mentioned in \citet
{SetF14} as to have an effect on LDL, while variants with names in red
are not [if a variant was not in dbSNP build 137, we named it by
indicating chromosome and position, following the convention in \citet
{SetF14}].}\label{fig:genetics2}
\end{figure}

Figure~\ref{fig:genetics2} focuses on the set of variants where there
is some disagreement between the 8 procedures we considered, after
eliminating the 90 variants selected only by the Lasso--$\LCV$. In
addition to recovering all except one of the variants identified in
\citet{SetF14}, and to the core of variants selected by all methods,
SLOPE--$\lambda_{\mathrm{G}^\star}$ selects 3 rare variants and 3
common variants. While the
rare variants were not singularly analyzed in the original study, they
are in the two regions where aggregate tests highlighted the role of
this type of variation. One is in \textit{ABCA1} and the other two are in
\textit{CETP},
and they are both nonsynonimous. Two of the three additional common
variants are in \textit{CETP} and one is in {\em MADD}; in addition to
SLOPE, these
are selected by Lasso--$\LCV$ and the marginal tests. One of the
common variants and one rare variant in {\em CETP} are mentioned as a result
of the limited foray in model selection in \citet{SetF14}. SLOPE--$\LMC$
selects two less of these variants.

In order to get a handle on the effective FDR control of SLOPE in this
setting, we resorted to simulations. We consider a number $k$ of
relevant variants ranging from 0 to 100, while concentrating on lower
values. At each level, $k$ columns of the design matrix were selected
at random and assigned an effect of $\sqrt{2\log p}$ against a noise
level $\sigma$ set to 1. While analyzing the data with $\LMC$ and
$\lambda_{\mathrm{G}^\star}$, we estimated $\sigma$ from the full
model in each run. Figure~\ref{fig:fdr_Chiara}(b)--(c) reports the
average FDP across 500
replicates and their standard error: the FDR of both $\LMC$ and
$\lambda_{\mathrm{G}^\star}$
are close to the nominal levels for all $k\leq100$.

In conclusion, the analysis with SLOPE confirms the results in
\citet{SetF14}, does not appear to introduce a large number of false
positives and, hence, makes it easier to include in the final list
of relevant variants a number of polymorphisms that are either
directly highlighted in the original paper or in regions that were
described as including a plurality of signals, but for which the
original multi-step analysis did not allow to make a
precise statement.

\section{Discussion}\label{sec4}
\label{sec:discussion}

The ease with which data are presently acquired has effectively
created a new scientific paradigm. In addition to carefully designing
experiments to test specific hypotheses, researchers often collect
data first, leaving question formulation to a later stage. In this
context, linear regression has increasingly been used to identify
connections between one response and a large number $p$ of possible
explanatory variables. When $p\gg n$, approaches based on convex
optimization have been particularly effective. An easily computable
solution has the advantage of definitiveness and of
reproducibility---another researcher, working on the same data set,
would obtain the same answer. Reproducibility of a scientific finding
or of the association between the outcome and the set of explanatory
variables selected among many, however, is harder to
achieve. Traditional tools such as $p$-values are often unhelpful in this
context because of the difficulties of accounting for the effect of
selection. In response, a great number of \mbox{proposals} [see,
e.g.,~\citet{BY05,POSI,B13,E11}, \citeauthor{JM13J} (\citeyear{JM13J,JM13Jb}), \citet{TibsTaylor,MB10,wasserman,Meinpvalues,Sara}, \citeauthor{zhangzhang}~(\citeyear{zhangzhang})]
present different
approaches for controlling some measures of type I error in the
context of variable selection. We here chose as a useful paradigm
that of controlling the expected proportion of irrelevant variables
among the selected ones. A similar goal of FDR control is pursued in
\citet{ko,max}. While \citet{ko} achieve exact FDR control in finite
sample irrespective of the structure of the design matrix, this
method, at least in the current implementation, is really best
tailored for cases where $n>p$. The work in \citet{max} relies on
$p$-values evaluated as in \citet{TibsTaylor}, and is limited to the
contexts where the assumptions in \citet{TibsTaylor} are met, including
the assumption that all true regressors appear before the false
regressors along the Lasso path. SLOPE controls FDR under orthogonal
designs, and simulation studies also show that SLOPE can keep the FDR
close to the nominal level when $p>n$ and the true model is sparse,
while offering large power and accurate prediction. This is, of
course, only a starting point and many open problems remain.

First, while our heuristics for the choice of the $\lambda$ sequence
allows to keep FDR under control for Gaussian designs and other random
design matrices [more examples are provided in \citet{SLOPE}], it is by
no means a definite solution. Further theoretical research is needed
to identify the sequences $\lambda$, which would provably control FDR
for these designs and other typical design matrices.

Second, just as in the BH procedure where the test statistics are
compared with fixed critical values, we have only considered in this
paper fixed values of the regularizing sequence $\{\lambda_i\}$. It
would be interesting to know whether it is possible to select such
parameters in a data-driven fashion as to achieve desirable
statistical properties. For the simpler Lasso problem, for instance, an
important question is whether it is possible to select $\lambda$ on
the Lasso path as to control the FDR. In the case where $n\geq p$, a
method to obtain this goal was recently proposed in \citet{ko}. It
would be of great interest to know if similar positive theoretical
results can be obtained for SLOPE, in perhaps restricted sparse
settings.

Third, our research points out the limits of signal sparsity which can be
handled by SLOPE. Such limitations are inherent to $\ell_1$ convex
optimization methods and also pertain to Lasso. Some discussion on
the minimal FDR which can be obtained with Lasso under Gaussian
designs is provided in \citet{SLOPE}, while new evocative results on
adaptive versions of Lasso are on the way.

Fourth, we illustrated the potential of SLOPE for multiple testing
with positively correlated test statistics. In our simple ANOVA
model, SLOPE controls FDR even when the unknown variance
components are replaced with their estimates. It remains an open
problem to theoretically describe a possibly larger class of
unknown covariance matrices for which SLOPE can be used
effectively.

In conclusion, we hope that the work presented so far would convince
the reader that SLOPE is an interesting convex program with promising
applications in statistics and motivates further research.\vspace*{-3pt}

\section*{Acknowledgments}
We would like to thank the Editor, Professor Karen Kafadar, the
Associate Editor and two reviewers for many constructive suggestions,
which led to a substantial improvement of this article.

We thank the
authors of \citet{SetF14} for letting us use their data during the
completion of dbGaP release. Emmanuel~J. Cand\`es would like to thank Stephen Becker
for all his help in integrating the sorted $\ell_1$ norm software into
TFOCS. Ma{\l}gorzata Bogdan would like to thank David Donoho and David Siegmund for
encouragement and Hatef Monajemi for helpful discussions. We are very
grateful to Lucas Janson for suggesting the acronym SLOPE, and to Rina
Foygel Barber and Julie Josse for useful comments about an early
version of the manuscript.\vspace*{-2pt}

\begin{supplement}[id=suppA]\vspace*{-2pt}
\stitle{Supplement to ``SLOPE---Adaptive variable selection via convex~optimization.''}
\slink[doi]{10.1214/15-AOAS842SUPP} 
\sdatatype{.pdf}
\sfilename{AOAS842\_supp.pdf}
\sdescription{The online Appendix contains  proofs of some technical results discussed in the text.}
\end{supplement}\vspace*{-2pt}

%

\printaddresses

\begin{thebibliography}{55}

\bibitem[\protect\citeauthoryear{Abramovich and
Benjamini}{1995}]{AbramovichBenjamini95}
%
\begin{bincollection}[auto:parserefs-M02]
\bauthor{\bsnm{Abramovich},~\bfnm{F.}\binits{F.}} \AND
\bauthor{\bsnm{Benjamini},~\bfnm{Y.}\binits{Y.}}
(\byear{1995}).
\btitle{Thresholding of wavelet coefficients as multiple hypotheses
testing procedure}.
In \bbooktitle{Wavelets and Statistics}.
\bseries{Lecture Notes in Statistics}
\bvolume{103}
\bpages{5--14}.
\bpublisher{Springer},
\blocation{Berlin}.
\end{bincollection}\vadjust{\goodbreak}
%

\bptok{imsref}%
\endbibitem

\bibitem[\protect\citeauthoryear{Abramovich et~al.}{2006}]{ABDJ}
%
\begin{barticle}[mr]
\bauthor{\bsnm{Abramovich},~\bfnm{Felix}\binits{F.}},
\bauthor{\bsnm{Benjamini},~\bfnm{Yoav}\binits{Y.}},
\bauthor{\bsnm{Donoho},~\bfnm{David~L.}\binits{D.~L.}} \AND
\bauthor{\bsnm{Johnstone},~\bfnm{Iain~M.}\binits{I.~M.}}
(\byear{2006}).
\btitle{Adapting to unknown sparsity by controlling the false
discovery rate}.
\bjournal{Ann. Statist.}
\bvolume{34}
\bpages{584--653}.
\bid{doi={10.1214/009053606000000074}, issn={0090-5364}, mr={2281879}}
\end{barticle}
%

\bptok{imsref}%
\endbibitem

\bibitem[\protect\citeauthoryear{Akaike}{1974}]{AIC}
%
\begin{barticle}[mr]
\bauthor{\bsnm{Akaike},~\bfnm{Hirotugu}\binits{H.}}
(\byear{1974}).
\btitle{A new look at the statistical model identification}.
\bjournal{IEEE Trans. Automat. Control}
\bvolume{AC-19}
\bpages{716--723}.
\bnote{System identification and time-series analysis}.
\bid{issn={0018-9286}, mr={0423716}}
\end{barticle}
%

\bptok{imsref}%
\endbibitem

\bibitem[\protect\citeauthoryear{Barlow et~al.}{1972}]{barlow72}
%
\begin{bbook}[mr]
\bauthor{\bsnm{Barlow},~\bfnm{R.~E.}\binits{R.~E.}},
\bauthor{\bsnm{Bartholomew},~\bfnm{D.~J.}\binits{D.~J.}},
\bauthor{\bsnm{Bremner},~\bfnm{J.~M.}\binits{J.~M.}} \AND
\bauthor{\bsnm{Brunk},~\bfnm{H.~D.}\binits{H.~D.}}
(\byear{1972}).
\btitle{Statistical Inference Under Order Restrictions. {T}he Theory
and Application of Isotonic Regression}.
\bpublisher{Wiley},
\blocation{New York}.
\bid{mr={0326887}}
\end{bbook}
%

\bptok{imsref}%
\endbibitem

\bibitem[\protect\citeauthoryear{Bauer, P{\"o}tscher and Hackl}{1988}]{Bauer}
%
\begin{barticle}[mr]
\bauthor{\bsnm{Bauer},~\bfnm{Peter}\binits{P.}},
\bauthor{\bsnm{P{\"o}tscher},~\bfnm{Benedikt~M.}\binits{B.~M.}} \AND
\bauthor{\bsnm{Hackl},~\bfnm{Peter}\binits{P.}}
(\byear{1988}).
\btitle{Model selection by multiple test procedures}.
\bjournal{Statistics}
\bvolume{19}
\bpages{39--44}.
\bid{doi={10.1080/02331888808802068}, issn={0233-1888}, mr={0921623}}
\end{barticle}
%

\bptok{imsref}%
\endbibitem

\bibitem[\protect\citeauthoryear{Beck and Teboulle}{2009}]{fista}
%
\begin{barticle}[mr]
\bauthor{\bsnm{Beck},~\bfnm{Amir}\binits{A.}} \AND
\bauthor{\bsnm{Teboulle},~\bfnm{Marc}\binits{M.}}
(\byear{2009}).
\btitle{A fast iterative shrinkage-thresholding algorithm for linear
inverse problems}.
\bjournal{SIAM J. Imaging Sci.}
\bvolume{2}
\bpages{183--202}.
\bid{doi={10.1137/080716542}, issn={1936-4954}, mr={2486527}}
\end{barticle}
%

\bptok{imsref}%
\endbibitem

\bibitem[\protect\citeauthoryear{Becker, Cand{\`e}s and Grant}{2011}]{tfocs}
%
\begin{barticle}[mr]
\bauthor{\bsnm{Becker},~\bfnm{Stephen~R.}\binits{S.~R.}},
\bauthor{\bsnm{Cand{\`e}s},~\bfnm{Emmanuel~J.}\binits{E.~J.}} \AND
\bauthor{\bsnm{Grant},~\bfnm{Michael~C.}\binits{M.~C.}}
(\byear{2011}).
\btitle{Templates for convex cone problems with applications to sparse
signal recovery}.
\bjournal{Math. Program. Comput.}
\bvolume{3}
\bpages{165--218}.
\bid{doi={10.1007/s12532-011-0029-5}, issn={1867-2949}, mr={2833262}}
\end{barticle}
%

\bptok{imsref}%
\endbibitem

\bibitem[\protect\citeauthoryear{Benjamini and Gavrilov}{2009}]{BG09}
%
\begin{barticle}[mr]
\bauthor{\bsnm{Benjamini},~\bfnm{Yoav}\binits{Y.}} \AND
\bauthor{\bsnm{Gavrilov},~\bfnm{Yulia}\binits{Y.}}
(\byear{2009}).
\btitle{A simple forward selection procedure based on false discovery
rate control}.
\bjournal{Ann. Appl. Stat.}
\bvolume{3}
\bpages{179--198}.
\bid{doi={10.1214/08-AOAS194}, issn={1932-6157}, mr={2668704}}
\end{barticle}
%

\bptok{imsref}%
\endbibitem

\bibitem[\protect\citeauthoryear{Benjamini and Hochberg}{1995}]{BH95}
%
\begin{barticle}[mr]
\bauthor{\bsnm{Benjamini},~\bfnm{Yoav}\binits{Y.}} \AND
\bauthor{\bsnm{Hochberg},~\bfnm{Yosef}\binits{Y.}}
(\byear{1995}).
\btitle{Controlling the false discovery rate: A practical and powerful
approach to multiple testing}.
\bjournal{J. Roy. Statist. Soc. Ser. B}
\bvolume{57}
\bpages{289--300}.
\bid{issn={0035-9246}, mr={1325392}}
\end{barticle}
%

\bptok{imsref}%
\endbibitem

\bibitem[\protect\citeauthoryear{Benjamini and Yekutieli}{2005}]{BY05}
%
\begin{barticle}[mr]
\bauthor{\bsnm{Benjamini},~\bfnm{Yoav}\binits{Y.}} \AND
\bauthor{\bsnm{Yekutieli},~\bfnm{Daniel}\binits{D.}}
(\byear{2005}).
\btitle{False discovery rate-adjusted multiple confidence intervals
for selected parameters}.
\bjournal{J. Amer. Statist. Assoc.}
\bvolume{100}
\bpages{71--93}.
\bid{doi={10.1198/016214504000001907}, issn={0162-1459}, mr={2156820}}
\end{barticle}
%

\bptok{imsref}%
\endbibitem

\bibitem[\protect\citeauthoryear{Berk et~al.}{2013}]{POSI}
%
\begin{barticle}[mr]
\bauthor{\bsnm{Berk},~\bfnm{Richard}\binits{R.}},
\bauthor{\bsnm{Brown},~\bfnm{Lawrence}\binits{L.}},
\bauthor{\bsnm{Buja},~\bfnm{Andreas}\binits{A.}},
\bauthor{\bsnm{Zhang},~\bfnm{Kai}\binits{K.}} \AND
\bauthor{\bsnm{Zhao},~\bfnm{Linda}\binits{L.}}
(\byear{2013}).
\btitle{Valid post-selection inference}.
\bjournal{Ann. Statist.}
\bvolume{41}
\bpages{802--837}.
\bid{doi={10.1214/12-AOS1077}, issn={0090-5364}, mr={3099122}}
\end{barticle}
%

\bptok{imsref}%
\endbibitem

\bibitem[\protect\citeauthoryear{Best and Chakravarti}{1990}]{best90}
%
\begin{barticle}[mr]
\bauthor{\bsnm{Best},~\bfnm{Michael~J.}\binits{M.~J.}} \AND
\bauthor{\bsnm{Chakravarti},~\bfnm{Nilotpal}\binits{N.}}
(\byear{1990}).
\btitle{Active set algorithms for isotonic regression; a unifying framework}.
\bjournal{Math. Program.}
\bvolume{47}
\bpages{425--439}.
\bid{doi={10.1007/BF01580873}, issn={0025-5610}, mr={1068274}}
\end{barticle}\vadjust{\goodbreak}
%

\bptok{imsref}%
\endbibitem

\bibitem[\protect\citeauthoryear{Birg{\'e} and Massart}{2001}]{BirgeMassart}
%
\begin{barticle}[mr]
\bauthor{\bsnm{Birg{\'e}},~\bfnm{Lucien}\binits{L.}} \AND
\bauthor{\bsnm{Massart},~\bfnm{Pascal}\binits{P.}}
(\byear{2001}).
\btitle{Gaussian model selection}.
\bjournal{J. Eur. Math. Soc. (JEMS)}
\bvolume{3}
\bpages{203--268}.
\bid{doi={10.1007/s100970100031}, issn={1435-9855}, mr={1848946}}
\end{barticle}
%

\bptok{imsref}%
\endbibitem



\bibitem[\protect\citeauthoryear{Bogdan et~al.}{2011}]{ABOS}
%
\begin{barticle}[mr]
\bauthor{\bsnm{Bogdan},~\bfnm{Ma{\l}gorzata}\binits{M.}},
\bauthor{\bsnm{Chakrabarti},~\bfnm{Arijit}\binits{A.}},
\bauthor{\bsnm{Frommlet},~\bfnm{Florian}\binits{F.}} \AND
\bauthor{\bsnm{Ghosh},~\bfnm{Jayanta~K.}\binits{J.~K.}}
(\byear{2011}).
\btitle{Asymptotic {B}ayes-optimality under sparsity of some multiple
testing procedures}.
\bjournal{Ann. Statist.}
\bvolume{39}
\bpages{1551--1579}.
\bid{doi={10.1214/10-AOS869}, issn={0090-5364}, mr={2850212}}
\end{barticle}
%

\bptok{imsref}%
\endbibitem

\bibitem[\protect\citeauthoryear{Bogdan, Ghosh and {\.{Z}}ak-Szatkowska}{2008}]{QREI}
%
\begin{barticle}[auto:parserefs-M02]
\bauthor{\bsnm{Bogdan},~\bfnm{M.}\binits{M.}},
\bauthor{\bsnm{Ghosh},~\bfnm{J.~K.}\binits{J.~K.}} \AND
\bauthor{\bsnm{{\.{Z}}ak-Szatkowska},~\bfnm{M.}\binits{M.}}
(\byear{2008}).
\btitle{Selecting explanatory variables with the modified version of
{B}ayesian information criterion}.
\bjournal{Qual. Reliab. Eng. Int.}
\bvolume{24}
\bpages{627--641}.
\end{barticle}
%

\bptok{imsref}%
\endbibitem

\bibitem[\protect\citeauthoryear{Bogdan et~al.}{2015}]{supple}
%
\begin{bmisc}[author]
\bauthor{\bsnm{Bogdan},~\binits{M.}},
\bauthor{\bsnm{van den Berg},~\binits{E.}},
\bauthor{\bsnm{Sabatti},~\binits{C.}},
\bauthor{\bsnm{Su},~\binits{W.}} \AND
\bauthor{\bsnm{Cand\`es},~\binits{E. J.}}
(\byear{2015}).
\bhowpublished{Supplement to ``SLOPE---Adaptive variable selection via
convex~optimization.''
DOI:\doiurl{10.1214/15-AOAS842SUPP}}.
\bptok{imsref}%
\end{bmisc}
%

\bptok{imsref}%
\endbibitem

\bibitem[\protect\citeauthoryear{Bogdan et~al.}{2013}]{SLOPE}
%
\begin{bmisc}[auto:parserefs-M02]
\bauthor{\bsnm{Bogdan},~\bfnm{M.}\binits{M.}},
\bauthor{\bsnm{van~den Berg},~\bfnm{E.}\binits{E.}},
\bauthor{\bsnm{Su},~\bfnm{W.}\binits{W.}} \AND
\bauthor{\bsnm{Cand\`es},~\bfnm{E.~J.}\binits{E.~J.}}
(\byear{2013}).
\bhowpublished{Statistical estimation and testing via the ordered
$\ell_1$ norm.
Preprint. Available at \arxivurl{arXiv:1310.1969v2}.}
\end{bmisc}
%

\bptok{imsref}%
\endbibitem



\bibitem[\protect\citeauthoryear{Bondell and Reich}{2008}]{OSCAR1}
%
\begin{barticle}[mr]
\bauthor{\bsnm{Bondell},~\bfnm{Howard~D.}\binits{H.~D.}} \AND
\bauthor{\bsnm{Reich},~\bfnm{Brian~J.}\binits{B.~J.}}
(\byear{2008}).
\btitle{Simultaneous regression shrinkage, variable selection, and
supervised clustering of predictors with {OSCAR}}.
\bjournal{Biometrics}
\bvolume{64}
\bpages{115--123, 322--323}.
\bid{doi={10.1111/j.1541-0420.2007.00843.x}, issn={0006-341X}, mr={2422825}}
\bptnote{check volume, check pages}%
\end{barticle}
%

\bptok{imsref}%
\endbibitem

\bibitem[\protect\citeauthoryear{B{\"u}hlmann}{2013}]{B13}
%
\begin{barticle}[mr]
\bauthor{\bsnm{B{\"u}hlmann},~\bfnm{Peter}\binits{P.}}
(\byear{2013}).
\btitle{Statistical significance in high-dimensional linear models}.
\bjournal{Bernoulli}
\bvolume{19}
\bpages{1212--1242}.
\bid{doi={10.3150/12-BEJSP11}, issn={1350-7265}, mr={3102549}}
\end{barticle}
%

\bptok{imsref}%
\endbibitem

\bibitem[\protect\citeauthoryear{Candes and Tao}{2007}]{DS}
%
\begin{barticle}[mr]
\bauthor{\bsnm{Candes},~\bfnm{Emmanuel}\binits{E.}} \AND
\bauthor{\bsnm{Tao},~\bfnm{Terence}\binits{T.}}
(\byear{2007}).
\btitle{The {D}antzig selector: Statistical estimation when {$p$} is
much larger than {$n$}}.
\bjournal{Ann. Statist.}
\bvolume{35}
\bpages{2313--2351}.
\bid{doi={10.1214/009053606000001523}, issn={0090-5364}, mr={2382644}}
\end{barticle}
%

\bptok{imsref}%
\endbibitem

\bibitem[\protect\citeauthoryear{Cand{\`e}s, Wakin and Boyd}{2008}]{relas}
%
\begin{barticle}[mr]
\bauthor{\bsnm{Cand{\`e}s},~\bfnm{Emmanuel~J.}\binits{E.~J.}},
\bauthor{\bsnm{Wakin},~\bfnm{Michael~B.}\binits{M.~B.}} \AND
\bauthor{\bsnm{Boyd},~\bfnm{Stephen~P.}\binits{S.~P.}}
(\byear{2008}).
\btitle{Enhancing sparsity by reweighted {$l\sb1$} minimization}.
\bjournal{J. Fourier Anal. Appl.}
\bvolume{14}
\bpages{877--905}.
\bid{doi={10.1007/s00041-008-9045-x}, issn={1069-5869}, mr={2461611}}
\end{barticle}
%

\bptok{imsref}%
\endbibitem

\bibitem[\protect\citeauthoryear{de~Leeuw, Hornik and
Mair}{2009}]{deleeuw2009isotone}
%
\begin{barticle}[auto:parserefs-M02]
\bauthor{\bparticle{de} \bsnm{Leeuw},~\bfnm{J.}\binits{J.}},
\bauthor{\bsnm{Hornik},~\bfnm{K.}\binits{K.}} \AND
\bauthor{\bsnm{Mair},~\bfnm{P.}\binits{P.}}
(\byear{2009}).
\btitle{Isotone optimization in R: Pool-adjacent-violators algorithm
(PAVA) and active set methods}.
\bjournal{J. Stat. Softw.}
\bvolume{32}
\bpages{1--24}.
\end{barticle}
%

\bptok{imsref}%
\endbibitem

\bibitem[\protect\citeauthoryear{Efron}{2011}]{E11}
%
\begin{barticle}[mr]
\bauthor{\bsnm{Efron},~\bfnm{Bradley}\binits{B.}}
(\byear{2011}).
\btitle{Tweedie's formula and selection bias}.
\bjournal{J. Amer. Statist. Assoc.}
\bvolume{106}
\bpages{1602--1614}.
\bid{doi={10.1198/jasa.2011.tm11181}, issn={0162-1459}, mr={2896860}}
\end{barticle}
%

\bptok{imsref}%
\endbibitem

\bibitem[\protect\citeauthoryear{Foster and George}{1994}]{RIC}
%
\begin{barticle}[mr]
\bauthor{\bsnm{Foster},~\bfnm{Dean~P.}\binits{D.~P.}} \AND
\bauthor{\bsnm{George},~\bfnm{Edward~I.}\binits{E.~I.}}
(\byear{1994}).
\btitle{The risk inflation criterion for multiple regression}.
\bjournal{Ann. Statist.}
\bvolume{22}
\bpages{1947--1975}.
\bid{doi={10.1214/aos/1176325766}, issn={0090-5364}, mr={1329177}}
\end{barticle}
%

\bptok{imsref}%
\endbibitem

\bibitem[\protect\citeauthoryear{Foster and Stine}{1999}]{FosterStine}
%
\begin{barticle}[mr]
\bauthor{\bsnm{Foster},~\bfnm{Dean~P.}\binits{D.~P.}} \AND
\bauthor{\bsnm{Stine},~\bfnm{Robert~A.}\binits{R.~A.}}
(\byear{1999}).
\btitle{Local asymptotic coding and the minimum description length}.
\bjournal{IEEE Trans. Inform. Theory}
\bvolume{45}
\bpages{1289--1293}.
\bid{doi={10.1109/18.761287}, issn={0018-9448}, mr={1686271}}
\end{barticle}
%

\bptok{imsref}%
\endbibitem

\bibitem[\protect\citeauthoryear{Foygel-Barber and Cand{\`{e}}s}{2014}]{ko}
%
\begin{barticle}[auto:parserefs-M02]
\bauthor{\bsnm{Foygel-Barber},~\bfnm{R.}\binits{R.}} \AND
\bauthor{\bsnm{Cand{\`{e}}s},~\bfnm{E.~J.}\binits{E.~J.}}
(\byear{2014}).
\btitle{Controlling the false discovery rate via knockoffs}.
\bjournal{Ann. Statist.}
\bnote{To appear. Available at \arxivurl{arXiv:1404.5609}.}
\end{barticle}
%

\bptok{imsref}%
\endbibitem

\bibitem[\protect\citeauthoryear{Frommlet and Bogdan}{2013}]{FB}
%
\begin{barticle}[mr]
\bauthor{\bsnm{Frommlet},~\bfnm{Florian}\binits{F.}} \AND
\bauthor{\bsnm{Bogdan},~\bfnm{Ma{\l}gorzata}\binits{M.}}
(\byear{2013}).
\btitle{Some optimality properties of FDR controlling rules under sparsity}.
\bjournal{Electron. J. Stat.}
\bvolume{7}
\bpages{1328--1368}.
\bid{doi={10.1214/13-EJS808}, issn={1935-7524}, mr={3063610}}
\end{barticle}
%

\bptok{imsref}%
\endbibitem

\bibitem[\protect\citeauthoryear{Frommlet et~al.}{2012}]{GWAS2012}
%
\begin{barticle}[mr]
\bauthor{\bsnm{Frommlet},~\bfnm{Florian}\binits{F.}},
\bauthor{\bsnm{Ruhaltinger},~\bfnm{Felix}\binits{F.}},
\bauthor{\bsnm{Twar{\'o}g},~\bfnm{Piotr}\binits{P.}} \AND
\bauthor{\bsnm{Bogdan},~\bfnm{Ma{\l}gorzata}\binits{M.}}
(\byear{2012}).
\btitle{Modified versions of {B}ayesian information criterion for
genome-wide association studies}.
\bjournal{Comput. Statist. Data Anal.}
\bvolume{56}
\bpages{1038--1051}.
\bid{doi={10.1016/j.csda.2011.05.005}, issn={0167-9473}, mr={2897552}}
\end{barticle}
%

\bptok{imsref}%
\endbibitem

\bibitem[\protect\citeauthoryear{Grazier~G'Sell, Hastie and
Tibshirani}{2013}]{max}
%
\begin{bmisc}[auto:parserefs-M02]
\bauthor{\bsnm{Grazier G'Sell},~\bfnm{M.}\binits{M.}},
\bauthor{\bsnm{Hastie},~\bfnm{T.}\binits{T.}} \AND
\bauthor{\bsnm{Tibshirani},~\bfnm{R.}\binits{R.}}
(\byear{2013}).
\bhowpublished{False variable selection rates in regression.
Preprint. Available at \arxivurl{arXiv:1302.2303}.}
\end{bmisc}
%

\bptok{imsref}%
\endbibitem

\bibitem[\protect\citeauthoryear{Grotzinger and Witzgall}{1984}]{odersimplex}
%
\begin{barticle}[mr]
\bauthor{\bsnm{Grotzinger},~\bfnm{S.~J.}\binits{S.~J.}} \AND
\bauthor{\bsnm{Witzgall},~\bfnm{C.}\binits{C.}}
(\byear{1984}).
\btitle{Projections onto order simplexes}.
\bjournal{Appl. Math. Optim.}
\bvolume{12}
\bpages{247--270}.
\bid{doi={10.1007/BF01449044}, issn={0095-4616}, mr={0768632}}
\end{barticle}
%

\bptok{imsref}%
\endbibitem

\bibitem[\protect\citeauthoryear{Ingster}{1998}]{ingster99}
%
\begin{barticle}[mr]
\bauthor{\bsnm{Ingster},~\bfnm{Yu.~I.}\binits{Yu.~I.}}
(\byear{1998}).
\btitle{Minimax detection of a signal for {$l\sp n$}-balls}.
\bjournal{Math. Methods Statist.}
\bvolume{7}
\bpages{401--428}.
\bid{issn={1066-5307}, mr={1680087}}
\bptnote{check pages, check year}%
\end{barticle}
%

\bptok{imsref}%
\endbibitem

\bibitem[\protect\citeauthoryear{Javanmard and Montanari}{2014a}]{JM13J}
%
\begin{barticle}[mr]
\bauthor{\bsnm{Javanmard},~\bfnm{Adel}\binits{A.}} \AND
\bauthor{\bsnm{Montanari},~\bfnm{Andrea}\binits{A.}}
(\byear{2014}a).
\btitle{Confidence intervals and hypothesis testing for
high-dimensional regression}.
\bjournal{J. Mach. Learn. Res.}
\bvolume{15}
\bpages{2869--2909}.
\bid{issn={1532-4435}, mr={3277152}}
\bptnote{check volume, check pages, check year}%
\end{barticle}
%

\bptok{imsref}%
\endbibitem

\bibitem[\protect\citeauthoryear{Javanmard and Montanari}{2014b}]{JM13Jb}
%
\begin{barticle}[mr]
\bauthor{\bsnm{Javanmard},~\bfnm{Adel}\binits{A.}} \AND
\bauthor{\bsnm{Montanari},~\bfnm{Andrea}\binits{A.}}
(\byear{2014}b).
\btitle{Hypothesis testing in high-dimensional regression under the
{G}aussian random design model: Asymptotic theory}.
\bjournal{IEEE Trans. Inform. Theory}
\bvolume{60}
\bpages{6522--6554}.
\bid{doi={10.1109/TIT.2014.2343629}, issn={0018-9448}, mr={3265038}}
\bptnote{check volume, check pages, check year}%
\end{barticle}
%

\bptok{imsref}%
\endbibitem

\bibitem[\protect\citeauthoryear{Kruskal}{1964}]{kruskal64}
%
\begin{barticle}[mr]
\bauthor{\bsnm{Kruskal},~\bfnm{J.~B.}\binits{J.~B.}}
(\byear{1964}).
\btitle{Nonmetric multidimensional scaling: A numerical method}.
\bjournal{Psychometrika}
\bvolume{29}
\bpages{115--129}.
\bid{issn={0033-3123}, mr={0169713}}
\end{barticle}
%

\bptok{imsref}%
\endbibitem

\bibitem[\protect\citeauthoryear{Lockhart et~al.}{2014}]{TibsTaylor}
%
\begin{barticle}[mr]
\bauthor{\bsnm{Lockhart},~\bfnm{Richard}\binits{R.}},
\bauthor{\bsnm{Taylor},~\bfnm{Jonathan}\binits{J.}},
\bauthor{\bsnm{Tibshirani},~\bfnm{Ryan~J.}\binits{R.~J.}} \AND
\bauthor{\bsnm{Tibshirani},~\bfnm{Robert}\binits{R.}}
(\byear{2014}).
\btitle{A significance test for the Lasso}.
\bjournal{Ann. Statist.}
\bvolume{42}
\bpages{413--468}.
\bid{doi={10.1214/13-AOS1175}, issn={0090-5364}, mr={3210970}}
\end{barticle}
%

\bptok{imsref}%
\endbibitem

\bibitem[\protect\citeauthoryear{Mallows}{1973}]{Cp}
%
\begin{barticle}[auto:parserefs-M02]
\bauthor{\bsnm{Mallows},~\bfnm{C.~L.}\binits{C.~L.}}
(\byear{1973}).
\btitle{Some comments on $c_p$}.
\bjournal{Technometrics}
\bvolume{15}
\bpages{661--676}.
\end{barticle}
%

\bptok{imsref}%
\endbibitem

\bibitem[\protect\citeauthoryear{Meinshausen}{2007}]{Meinrelaxed}
%
\begin{barticle}[mr]
\bauthor{\bsnm{Meinshausen},~\bfnm{Nicolai}\binits{N.}}
(\byear{2007}).
\btitle{Relaxed {L}asso}.
\bjournal{Comput. Statist. Data Anal.}
\bvolume{52}
\bpages{374--393}.
\bid{doi={10.1016/j.csda.2006.12.019}, issn={0167-9473}, mr={2409990}}
\end{barticle}
%

\bptok{imsref}%
\endbibitem

\bibitem[\protect\citeauthoryear{Meinshausen and B{\"u}hlmann}{2010}]{MB10}
%
\begin{barticle}[mr]
\bauthor{\bsnm{Meinshausen},~\bfnm{Nicolai}\binits{N.}} \AND
\bauthor{\bsnm{B{\"u}hlmann},~\bfnm{Peter}\binits{P.}}
(\byear{2010}).
\btitle{Stability selection}.
\bjournal{J. R. Stat. Soc. Ser. B. Stat. Methodol.}
\bvolume{72}
\bpages{417--473}.
\bid{doi={10.1111/j.1467-9868.2010.00740.x}, issn={1369-7412}, mr={2758523}}
\end{barticle}
%

\bptok{imsref}%
\endbibitem

\bibitem[\protect\citeauthoryear{Meinshausen, Meier and B{\"
u}hlmann}{2009}]{Meinpvalues}
%
\begin{barticle}[mr]
\bauthor{\bsnm{Meinshausen},~\bfnm{Nicolai}\binits{N.}},
\bauthor{\bsnm{Meier},~\bfnm{Lukas}\binits{L.}} \AND
\bauthor{\bsnm{B{\"u}hlmann},~\bfnm{Peter}\binits{P.}}
(\byear{2009}).
\btitle{{$p$}-values for high-dimensional regression}.
\bjournal{J. Amer. Statist. Assoc.}
\bvolume{104}
\bpages{1671--1681}.
\bid{doi={10.1198/jasa.2009.tm08647}, issn={0162-1459}, mr={2750584}}
\end{barticle}
%

\bptok{imsref}%
\endbibitem

\bibitem[\protect\citeauthoryear{Nesterov}{2004}]{Nesbook}
%
\begin{bbook}[mr]
\bauthor{\bsnm{Nesterov},~\bfnm{Yurii}\binits{Y.}}
(\byear{2004}).
\btitle{Introductory Lectures on Convex Optimization. A Basic Course}.
\bpublisher{Kluwer Academic},
\blocation{Boston, MA}.
\bid{doi={10.1007/978-1-4419-8853-9}, mr={2142598}}
\end{bbook}
%

\bptok{imsref}%
\endbibitem

\bibitem[\protect\citeauthoryear{Nesterov}{2007}]{nes07}
%
\begin{bmisc}[auto:parserefs-M02]
\bauthor{\bsnm{Nesterov},~\bfnm{Y.}\binits{Y.}}
(\byear{2007}).
\bhowpublished{Gradient methods for minimizing composite objective function.
CORE discussion paper. Center for Operations Research and Econometrics (CORE),
Universit\'e Catholique de Louvain.
Available at \surl{http://www.ecore.be/DPs/dp\_1191313936.pdf}.}
\end{bmisc}
%

\bptok{imsref}%
\endbibitem

\bibitem[\protect\citeauthoryear{Parikh and Boyd}{2013}]{BoydProx}
%
\begin{bincollection}[auto:parserefs-M02]
\bauthor{\bsnm{Parikh},~\bfnm{N.}\binits{N.}} \AND
\bauthor{\bsnm{Boyd},~\bfnm{S.}\binits{S.}}
(\byear{2013}).
\btitle{Proximal algorithms}.
In \bbooktitle{Foundations and Trends in Optimization}
\bvolume{1}
\bpages{123--231}.
\end{bincollection}
%

\bptok{imsref}%
\endbibitem

\bibitem[\protect\citeauthoryear{Sarkar}{2002}]{sarkar2002}
%
\begin{barticle}[mr]
\bauthor{\bsnm{Sarkar},~\bfnm{Sanat~K.}\binits{S.~K.}}
(\byear{2002}).
\btitle{Some results on false discovery rate in stepwise multiple
testing procedures}.
\bjournal{Ann. Statist.}
\bvolume{30}
\bpages{239--257}.
\bid{doi={10.1214/aos/1015362192}, issn={0090-5364}, mr={1892663}}
\bptnote{check volume}%
\end{barticle}
%

\bptok{imsref}%
\endbibitem

\bibitem[\protect\citeauthoryear{Service et~al.}{2014}]{SetF14}
%
\begin{barticle}[auto:parserefs-M02]
\bauthor{\bsnm{Service},~\bfnm{S.~K.}\binits{S.~K.}},
\bauthor{\bsnm{Teslovich},~\bfnm{T.~M.}\binits{T.~M.}},
\bauthor{\bsnm{Fuchsberger},~\bfnm{C.}\binits{C.}},
\bauthor{\bsnm{Ramensky},~\bfnm{V.}\binits{V.}},
\bauthor{\bsnm{Yajnik},~\bfnm{P.}\binits{P.}},
\bauthor{\bsnm{Koboldt},~\bfnm{D.~C.}\binits{D.~C.}},
\bauthor{\bsnm{Larson},~\bfnm{D.~E.}\binits{D.~E.}},
\bauthor{\bsnm{Zhang},~\bfnm{Q.}\binits{Q.}},
\bauthor{\bsnm{Lin},~\bfnm{L.}\binits{L.}},
\bauthor{\bsnm{Welch},~\bfnm{R.}\binits{R.}},
\bauthor{\bsnm{Ding},~\bfnm{L.}\binits{L.}},
\bauthor{\bsnm{McLellan},~\bfnm{M.~D.}\binits{M.~D.}},
\bauthor{\bsnm{O'Laughlin},~\bfnm{M.}\binits{M.}},
\bauthor{\bsnm{Fronick},~\bfnm{C.}\binits{C.}},
\bauthor{\bsnm{Fulton},~\bfnm{L.~L.}\binits{L.~L.}},
\bauthor{\bsnm{Magrini},~\bfnm{V.}\binits{V.}},
\bauthor{\bsnm{Swift},~\bfnm{A.}\binits{A.}},
\bauthor{\bsnm{Elliott},~\bfnm{P.}\binits{P.}},
\bauthor{\bsnm{Jarvelin},~\bfnm{M.~R.}\binits{M.~R.}},
\bauthor{\bsnm{Kaakinen},~\bfnm{M.}\binits{M.}},
\bauthor{\bsnm{McCarthy},~\bfnm{M.~I.}\binits{M.~I.}},
\bauthor{\bsnm{Peltonen},~\bfnm{L.}\binits{L.}},
\bauthor{\bsnm{Pouta},~\bfnm{A.}\binits{A.}},
\bauthor{\bsnm{Bonnycastle},~\bfnm{L.~L.}\binits{L.~L.}},
\bauthor{\bsnm{Collins},~\bfnm{F.~S.}\binits{F.~S.}},
\bauthor{\bsnm{Narisu},~\bfnm{N.}\binits{N.}},
\bauthor{\bsnm{Stringham},~\bfnm{H.~M.}\binits{H.~M.}},
\bauthor{\bsnm{Tuomilehto},~\bfnm{J.}\binits{J.}},
\bauthor{\bsnm{Ripatti},~\bfnm{S.}\binits{S.}},
\bauthor{\bsnm{Fulton},~\bfnm{R.~S.}\binits{R.~S.}},
\bauthor{\bsnm{Sabatti},~\bfnm{C.}\binits{C.}},
\bauthor{\bsnm{Wilson},~\bfnm{R.~K.}\binits{R.~K.}},
\bauthor{\bsnm{Boehnke},~\bfnm{M.}\binits{M.}} \AND
\bauthor{\bsnm{Freimer},~\bfnm{N.~B.}\binits{N.~B.}}
(\byear{2014}).
\btitle{{R}e-sequencing expands our understanding of the phenotypic
impact of variants at G{W}{A}{S} loci}.
\bjournal{PLoS Genet.}
\bvolume{10}
\bpages{e1004147}.
\end{barticle}
%

\bptok{imsref}%
\endbibitem

\bibitem[\protect\citeauthoryear{St{\"a}dler, B{\"u}hlmann and
van~de Geer}{2010}]{ScaledLasso}
%
\begin{barticle}[mr]
\bauthor{\bsnm{St{\"a}dler},~\bfnm{Nicolas}\binits{N.}},
\bauthor{\bsnm{B{\"u}hlmann},~\bfnm{Peter}\binits{P.}} \AND
\bauthor{\bsnm{van~de Geer},~\bfnm{Sara}\binits{S.}}
(\byear{2010}).
\btitle{{$\ell\sb1$}-penalization for mixture regression models}.
\bjournal{TEST}
\bvolume{19}
\bpages{209--256}.
\bid{doi={10.1007/s11749-010-0197-z}, issn={1133-0686}, mr={2677722}}
\bptnote{check related, check pages}%
\end{barticle}
%

\bptok{imsref}%
\endbibitem

\bibitem[\protect\citeauthoryear{Sun and Zhang}{2012}]{sun2012scaled}
%
\begin{barticle}[mr]
\bauthor{\bsnm{Sun},~\bfnm{Tingni}\binits{T.}} \AND
\bauthor{\bsnm{Zhang},~\bfnm{Cun-Hui}\binits{C.-H.}}
(\byear{2012}).
\btitle{Scaled sparse linear regression}.
\bjournal{Biometrika}
\bvolume{99}
\bpages{879--898}.
\bid{doi={10.1093/biomet/ass043}, issn={0006-3444}, mr={2999166}}
\end{barticle}
%

\bptok{imsref}%
\endbibitem

\bibitem[\protect\citeauthoryear{Tibshirani}{1996}]{Tibs96}
%
\begin{barticle}[mr]
\bauthor{\bsnm{Tibshirani},~\bfnm{Robert}\binits{R.}}
(\byear{1996}).
\btitle{Regression shrinkage and selection via the Lasso}.
\bjournal{J. Roy. Statist. Soc. Ser. B}
\bvolume{58}
\bpages{267--288}.
\bid{issn={0035-9246}, mr={1379242}}
\end{barticle}
%

\bptok{imsref}%
\endbibitem

\bibitem[\protect\citeauthoryear{Tibshirani and
Knight}{1999}]{TibshiraniKnight}
%
\begin{barticle}[mr]
\bauthor{\bsnm{Tibshirani},~\bfnm{Robert}\binits{R.}} \AND
\bauthor{\bsnm{Knight},~\bfnm{Keith}\binits{K.}}
(\byear{1999}).
\btitle{The covariance inflation criterion for adaptive model selection}.
\bjournal{J. R. Stat. Soc. Ser. B. Stat. Methodol.}
\bvolume{61}
\bpages{529--546}.
\bid{doi={10.1111/1467-9868.00191}, issn={1369-7412}, mr={1707859}}
\bptnote{check volume, check pages}%
\end{barticle}
%

\bptok{imsref}%
\endbibitem

\bibitem[\protect\citeauthoryear{van~de Geer et~al.}{2014}]{Sara}
%
\begin{barticle}[mr]
\bauthor{\bsnm{van~de Geer},~\bfnm{Sara}\binits{S.}},
\bauthor{\bsnm{B{\"u}hlmann},~\bfnm{Peter}\binits{P.}},
\bauthor{\bsnm{Ritov},~\bfnm{Ya'acov}\binits{Y.}} \AND
\bauthor{\bsnm{Dezeure},~\bfnm{Ruben}\binits{R.}}
(\byear{2014}).
\btitle{On asymptotically optimal confidence regions and tests for
high-dimensional models}.
\bjournal{Ann. Statist.}
\bvolume{42}
\bpages{1166--1202}.
\bid{doi={10.1214/14-AOS1221}, issn={0090-5364}, mr={3224285}}
\end{barticle}
%

\bptok{imsref}%
\endbibitem

\bibitem[\protect\citeauthoryear{Wasserman and Roeder}{2009}]{wasserman}
%
\begin{barticle}[mr]
\bauthor{\bsnm{Wasserman},~\bfnm{Larry}\binits{L.}} \AND
\bauthor{\bsnm{Roeder},~\bfnm{Kathryn}\binits{K.}}
(\byear{2009}).
\btitle{High-dimensional variable selection}.
\bjournal{Ann. Statist.}
\bvolume{37}
\bpages{2178--2201}.
\bid{doi={10.1214/08-AOS646}, issn={0090-5364}, mr={2543689}}
\bptnote{check volume}%
\end{barticle}
%

\bptok{imsref}%
\endbibitem

\bibitem[\protect\citeauthoryear{Wu and Zhou}{2013}]{wuzhou}
%
\begin{barticle}[mr]
\bauthor{\bsnm{Wu},~\bfnm{Zheyang}\binits{Z.}} \AND
\bauthor{\bsnm{Zhou},~\bfnm{Harrison~H.}\binits{H.~H.}}
(\byear{2013}).
\btitle{Model selection and sharp asymptotic minimaxity}.
\bjournal{Probab. Theory Related Fields}
\bvolume{156}
\bpages{165--191}.
\bid{doi={10.1007/s00440-012-0424-5}, issn={0178-8051}, mr={3055256}}
\end{barticle}
%

\bptok{imsref}%
\endbibitem

\bibitem[\protect\citeauthoryear{Zeng and Figueiredo}{2014}]{WSL1}
%
\begin{barticle}[auto:parserefs-M02]
\bauthor{\bsnm{Zeng},~\bfnm{X.}\binits{X.}} \AND
\bauthor{\bsnm{Figueiredo},~\bfnm{M.}\binits{M.}}
(\byear{2014}).
\btitle{Decreasing weighted sorted l1 regularization}.
\bjournal{IEEE Signal Process. Lett.}
\bpages{1240--1244}.
\end{barticle}
%

\bptok{imsref}%
\endbibitem

\bibitem[\protect\citeauthoryear{Zhang and Zhang}{2014}]{zhangzhang}
%
\begin{barticle}[mr]
\bauthor{\bsnm{Zhang},~\bfnm{Cun-Hui}\binits{C.-H.}} \AND
\bauthor{\bsnm{Zhang},~\bfnm{Stephanie~S.}\binits{S.~S.}}
(\byear{2014}).
\btitle{Confidence intervals for low dimensional parameters in high
dimensional linear models}.
\bjournal{J. R. Stat. Soc. Ser. B. Stat. Methodol.}
\bvolume{76}
\bpages{217--242}.
\bid{doi={10.1111/rssb.12026}, issn={1369-7412}, mr={3153940}}
\bptnote{check volume, check year}%
\end{barticle}
%

\bptok{imsref}%
\endbibitem

\bibitem[\protect\citeauthoryear{Zhong and Kwok}{2012}]{OSCAR}
%
\begin{barticle}[auto:parserefs-M02]
\bauthor{\bsnm{Zhong},~\bfnm{L.}\binits{L.}} \AND
\bauthor{\bsnm{Kwok},~\bfnm{J.}\binits{J.}}
(\byear{2012}).
\btitle{Efficient sparse modeling with automatic feature grouping}.
\bjournal{IEEE Trans. Neural Netw. Learn. Syst.}
\bpages{1436--1447}.
\end{barticle}
%

\bptok{imsref}%
\endbibitem

\bibitem[\protect\citeauthoryear{Zou}{2006}]{adlas}
%
\begin{barticle}[mr]
\bauthor{\bsnm{Zou},~\bfnm{Hui}\binits{H.}}
(\byear{2006}).
\btitle{The adaptive Lasso and its oracle properties}.
\bjournal{J. Amer. Statist. Assoc.}
\bvolume{101}
\bpages{1418--1429}.
\bid{doi={10.1198/016214506000000735}, issn={0162-1459}, mr={2279469}}
\end{barticle}
%

\bptok{imsref}%
\endbibitem

\end{thebibliography}
\end{document}